\documentclass[
 preprint,
 amsmath,amssymb,
 aps,%pra,
]{revtex4}

\usepackage{graphicx}% Include figure files
\usepackage{dcolumn}% Align table columns on decimal point
\usepackage{bm}% bold math
\usepackage{verbatim}
\usepackage[caption=false]{subfig}
\usepackage{floatrow}
\usepackage{color}
\usepackage[dvipsnames]{xcolor}

\newcommand{\bra}[1]{\ensuremath{\langle#1|}}
\newcommand{\ket}[1]{\ensuremath{|#1\rangle}}
\newcommand{\EEE}{\mathcal{E}}
\newcommand{\MMM}{\mathcal{M}}
\newcommand{\NNN}{\mathcal{N}}
\newcommand{\CCC}{\mathcal{C}}
\newcommand{\DDD}{\boldsymbol{\mathcal{D}}}

\newcommand{\aaa}{\textbf{a}}
\newcommand{\ccc}{\textbf{c}}

\newcommand{\bbb}{\textbf{b}}
\newcommand{\hhh}{\textbf{H}}
\newcommand{\ddd}{\textbf{D}}
\newcommand{\sss}{\boldsymbol{\sigma}}
\newcommand{\rrr}{\boldsymbol{\rho}}

\newcommand{\bmaj}{\textbf{B}}
\newcommand{\rmaj}{\textbf{R}}
\newcommand{\cmaj}{\textbf{C}}
\newcommand{\lmaj}{\textbf{L}}
\newcommand{\amaj}{\textbf{A}}
\newcommand{\proj}{\boldsymbol{\Pi}}
\newcommand{\III}{\textbf{I}}
\newcommand{\LLL}{\mathcal{L}}
\newcommand{\av}[1]{\ensuremath{\langle#1 \rangle}}

\begin{document}

%\setcounter{Maxaffil}{2}
%\preprint{APS/123-QED}

\title{Degeneracy-preserving quantum non-demolition measurement of parity-type observables for cat-qubits}% Force line breaks with \\

\author{Joachim Cohen$^{1}$}
 \email{joachim.cohen@inria.fr}
 \author{W. Clarke Smith$^{2}$}
 \email{clarke.smith@yale.edu}
 \author{Michel H. Devoret$^{2}$}
 \email{michel.devoret@yale.edu}
\author{Mazyar Mirrahimi$^{1,3}$}%
 \email{mazyar.mirrahimi@inria.fr}
\affiliation{%
 $^{1}$QUANTIC project-team, INRIA Paris, France\\
 $^{2}$Department of Applied Physics, Yale University, New Haven, CT 06520, USA\\
 $^{3}$Yale Quantum Institute, Yale University, New Haven, CT 06520, USA\\
}%


\begin{abstract}
A central requirement for any quantum error correction scheme is the ability to perform quantum non-demolition measurements of an error syndrome, corresponding to a special symmetry property of the encoding scheme. It is in particular important that such a measurement does not introduce extra error mechanisms, not included in the error model of the correction scheme. In this letter, we ensure such a robustness by designing an interaction with a measurement device that preserves the degeneracy of the measured observable. More precisely, we propose a scheme to perform continuous and quantum non-demolition measurement of photon-number parity in a microwave cavity. This corresponds to the error syndrome in a  class of error correcting codes called the cat-codes, which have recently proven to be efficient and versatile for quantum information processing. In our design, we exploit the strongly nonlinear Hamiltonian of a high-impedance Josephson circuit, coupling a high-Q cavity storage cavity mode to a low-Q readout one. By driving the readout resonator at its resonance, the phase of the reflected/transmitted signal carries directly exploitable information on parity-type observables for encoded cat-qubits of the high-Q mode. 
\end{abstract}

\maketitle

By encoding a qubit in a superposition of coherent states of a harmonic oscillator, one benefits from the redundancy provided by the infinite dimensional Hilbert space of the system to realize a quantum error correction (QEC) protocol. In a  set of theoretical and experimental results, various aspects of  encoding~\cite{Leghtas-al-PRA-2013,vlastakis-science-2013}, manipulation~\cite{Mirrahimi-al-NJP-2014,Albert-PRL-2016,Heeres-PRL-2015,Wang-Gao-Science-2016}, error syndrome measurement~\cite{sun-nature-2013} and full quantum error correction~\cite{Leghtas-al-PRL-2013,Ofek-Petrenko-Nature-2016} with these states have been explored. Most spectacularly, a recent experiment~\cite{Ofek-Petrenko-Nature-2016} demonstrated an enhancement of the error-corrected cat-code's lifetime with respect to all system components. The performance of the error correction is however limited by uncorrected error channels such as deterministic relaxation of the coherent states amplitude, dephasing induced by cavity's inherited anharmonicity, and most significantly the propagating errors from the ancillary transmon~\cite{Koch-et-al-07} used for error syndrome measurements.  

In an effort towards a fault-tolerant and scalable architecture for quantum information processing, we recently proposed a framework based on non-linear drives and dissipations to dynamically protect a degenerate manifold spanned by two or four coherent states against some of these error channels~\cite{Mirrahimi-al-NJP-2014}. Indeed, by engineering a non-linear coupling to a driven bath where the exchange of photons occurs mainly in pairs (or quadruples) of photons, one can stabilize a manifold spanned by two (resp. four) coherent states $\MMM_{2,\alpha}=\text{span}\{\ket{\pm\alpha}\}$ (resp. $\MMM_{4,\alpha}=\text{span}\{\ket{\pm\alpha},\ket{\pm i\alpha}\}$). {This stabilization suppresses, exponentially in $|\alpha|^2$, the phase-flip errors of a logical qubit} given  by $\ket{0}_L=\ket{\CCC_{\alpha}^+}, \ket{1}_L=\ket{\CCC_{\alpha}^-}$ (resp. $\ket{0}_L=\ket{\CCC_{\alpha}^{(0\text{mod}4)}}, \ket{1}_L=\ket{\CCC_{\alpha}^{(2\text{mod}4)}}$) where\small
\begin{align*}
&\ket{\CCC_{\alpha}^\pm}=\NNN_\pm(\ket{\alpha}\pm\ket{-\alpha}),\\ 
&\ket{\CCC_{\alpha}^{(0\text{mod}4)}}=\NNN_0(\ket{\CCC_{\alpha}^+}+\ket{\CCC_{i\alpha}^+}),~
 \ket{\CCC_{\alpha}^{(2\text{mod}4)}}=\NNN_2(\ket{\CCC_{\alpha}^+}-\ket{\CCC_{i\alpha}^+}),\\
&\ket{\CCC_{\alpha}^{(1\text{mod}4)}}=\NNN_1(\ket{\CCC_{\alpha}^-}-i\ket{\CCC_{i\alpha}^-}),~
 \ket{\CCC_{\alpha}^{(3\text{mod}4)}}=\NNN_3(\ket{\CCC_{\alpha}^-}+i\ket{\CCC_{i\alpha}^-}),
\end{align*}\normalsize
and $\NNN_{\pm},\NNN_{0,1,2,3}$ are normalization constants near $1/\sqrt{2}$. One therefore deals with logical qubits that are only susceptible to bit-flip errors, but on which one can perform a universal set of logical gates (see~\cite{Mirrahimi-al-NJP-2014,Albert-PRL-2016}). Such bit-flip errors can next be suppressed to first order by photon-number parity measurements as in~\cite{Ofek-Petrenko-Nature-2016}. Also, one can achieve higher-order correction through a register of such logical qubits and performing joint parity measurements between adjacent ones.

While initial experiments with two-photon driven dissipation~\cite{Leghtas-Science-2015} illustrate the viability of such a framework, many theoretical and experimental improvements are required in order to achieve a fully fault-tolerant architecture. One very important improvement concerns the quantum non-demolition (QND) measurement protocols. Indeed, a central requirement for all above proposals is the ability to measure observables such as photon number parity of a cavity mode, or joint parity of two cavity modes. Such single-mode or two-mode photon number parity measurements have been performed using an ancillary transmon and a Ramsey interferometry type scheme~\cite{brune-et-al:PhRevA92,sun-nature-2013,Wang-Gao-Science-2016}. They however suffer from an important degree of non fault-tolerance and represent the main limitation in QEC~\cite{Ofek-Petrenko-Nature-2016}. In this letter, we propose a new framework to perform QND measurement of various important parity-type  observables which could be integrated in a fault-tolerant architecture.    

The current measurement schemes~\cite{sun-nature-2013,Wang-Gao-Science-2016} are based on a dispersive coupling of the cavity mode to a transmon through a Hamiltonian of the form $-\hbar\chi \ket{e}\bra{e} \aaa^\dag \aaa$. The parity measurement is performed by initializing the transmon in the superposition $(\ket{g}+\ket{e})/\sqrt{2}$ and waiting for a time $\pi/\chi$. The $\ket{e}$ state of transmon will therefore acquire a $\pi$ phase only for odd cavity Fock states. A measurement of the transmon, distinguishing between $(\ket{g}+\ket{e})/\sqrt{2}$ and $(\ket{g}-\ket{e})/\sqrt{2}$  indicates the photon number parity. Nevertheless, a $T_1$ error of the transmon during the evolution propagates to the cavity mode inducing photon dephasing. Indeed, such a measurement protocol is not fault-tolerant as  the eigenstates of the measured observable (here parity cat states) get entangled to the ancillary system during the measurement protocol, making them vulnerable to the ancilla's errors (here $T_1$ errors):  a cat state $\ket{\CCC_{\alpha}^\pm}$ evolves to $(\ket{\CCC_{\alpha}^\pm}\otimes\ket{g}+\ket{\CCC_{\alpha e^{-i\chi t}}^\pm}\otimes\ket{e})/\sqrt{2}$. A fault-tolerant parity measurement could be for instance achieved through an effective Hamiltonian of the form $\hbar\chi \ket{e}\bra{e}\cos(\pi\aaa^\dag\aaa)$. A cat state $\ket{\CCC_{\alpha}^\pm}$ would then evolve to $\ket{\CCC_{\alpha}^\pm}\otimes(\ket{g}+e^{\pm i\chi t}\ket{e})/\sqrt{2}$, without entangling to the transmon. 

While the engineering of a highly degenerate Hamiltonian of the form $\hbar\chi\cos(\pi\aaa^\dag\aaa)$ seems to be a complicated task, we show  that in presence of two-photon or four-photon driven dissipation, it could be effectively achieved with the help of quantum Zeno dynamics~\cite{facchi-pascazio-2002}. By confining the dynamics to the manifold $\MMM_{2,\alpha}$, a physical Hamiltonian $\hhh$ acts as a projected one $\hhh_{\MMM_{2,\alpha}}=\Pi_{\MMM_{2,\alpha}} \hhh \Pi_{\MMM_{2,\alpha}}$, where $\Pi_{\MMM_{2,\alpha}}$ represents the projector on $\MMM_{2,\alpha}$.  To achieve an effective parity Hamiltonian, one requires a physical Hamiltonian $\hhh$ satisfying
$$
\hhh_{\MMM_{2,\alpha}}=\hbar\chi\Pi_{\MMM_{2,\alpha}}\cos(\pi\aaa^\dag\aaa)\Pi_{\MMM_{2,\alpha}}=\hbar\chi\sss_z^L,
$$
where $\sss_z^L$ is the Pauli operator along the $z$-axis of the logical qubit defined by $\{\ket{\CCC_{\alpha}^\pm}\}$ and well-approximated by $\ket{\alpha}\bra{-\alpha}+\ket{-\alpha}\bra{\alpha}$.
This means that $\hhh$ should couple the two coherent states $\ket{\pm\alpha}$. In the context of quantum superconducting circuits, such a Hamiltonian can be achieved by strongly coupling a high impedance cavity mode to a Josephson junction~\cite{Masluk-Devoret-PRL-2012,Pop-Devoret-Nature-2014}. Indeed, considering a cavity mode with frequency $\omega_a$ coupled capacitively to a Josephson junction, and assuming that  other modes (including the junction mode) are never excited, the effective Hamiltonian in the interaction picture will be of the form
$$
\hhh_{\text{int}}(t) = -\frac{E_J}{2}(\ddd[\beta(t)]+\ddd^{\dag}[\beta(t)]),\quad \beta(t) = i\varphi_a e^{i\omega_a t}.
$$ 
Here $E_J$  is the effective Josephson energy and ${\varphi_a= \sqrt{{Z_a}/{2R_Q}}}$, where $Z_a$ is the impedance of the cavity mode seen by the junction and $R_Q=(2e)^2/\hbar$ is the superconducting resistance quantum. Moreover, $\ddd[\beta(t)]$ is the displacement operator defined by $\ddd[\beta(t)] = e^{\beta(t)\aaa^\dag-\beta(t)^*\aaa}$. For $\varphi_a\approx 2|\alpha|$, this Hamiltonian couples the two coherent states $\ket{\pm\alpha}$. While a practical realization of such a high impedance cavity mode is discussed later, we provide here a  precise analysis of the effective Hamiltonian. 

\begin{figure}[!tbp]
  \centering
  \subfloat[]{\includegraphics[width=0.37\textwidth]{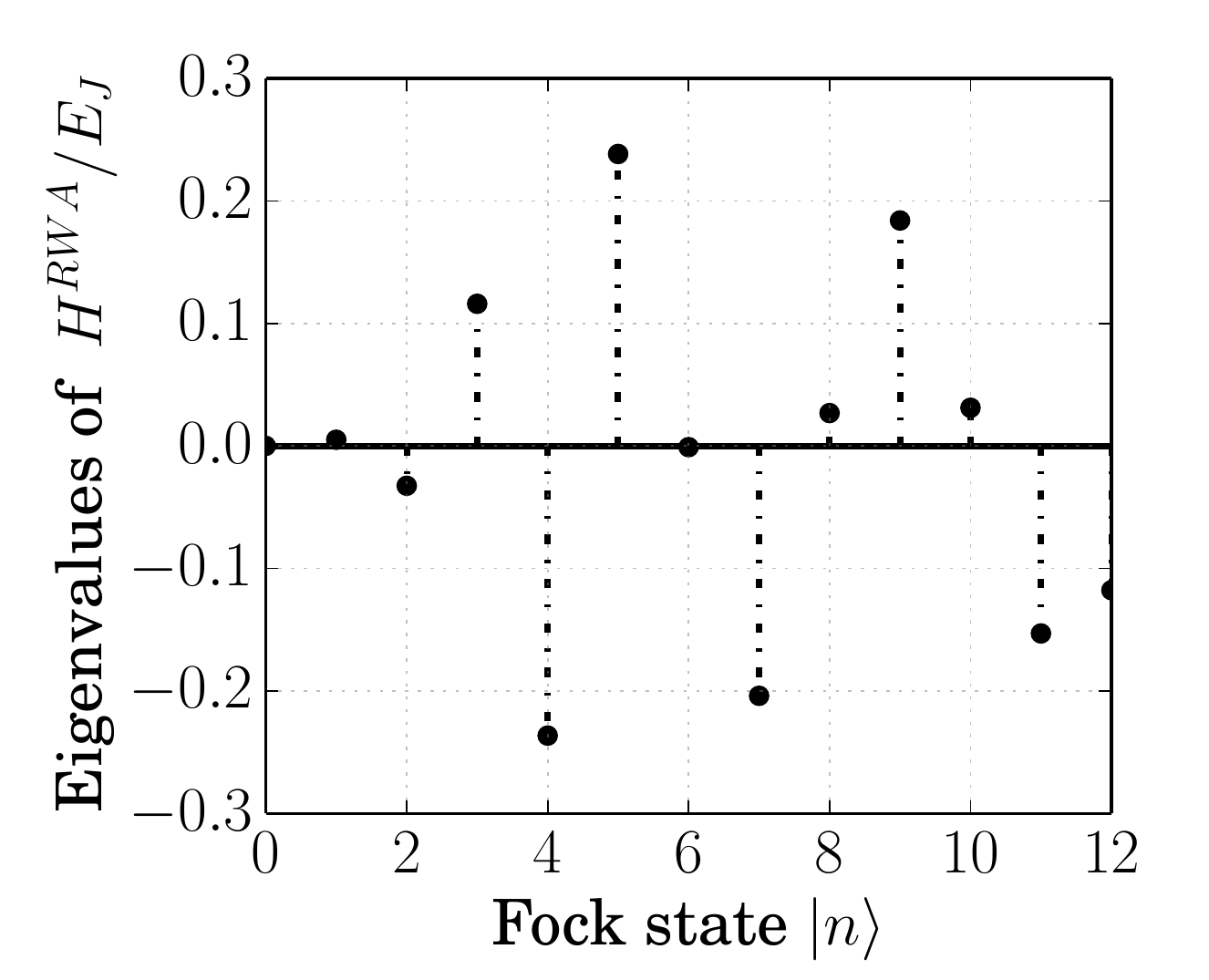}\label{fig:eigenvalues}}
\hfill
\subfloat[]{\includegraphics[width=0.37\textwidth]{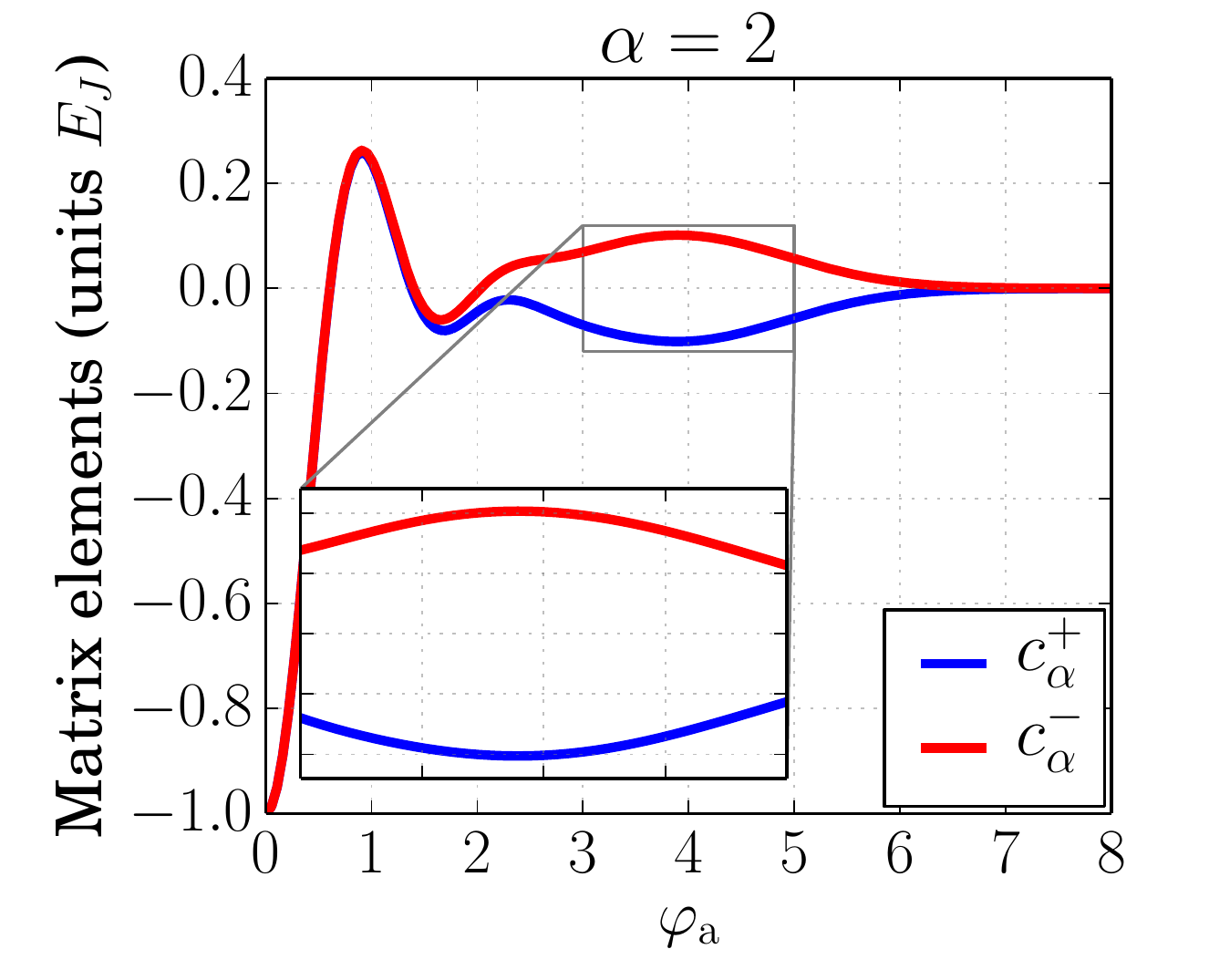}\label{fig:matrixelements}}
\hfill
\subfloat[]{\includegraphics[width=0.37\textwidth]{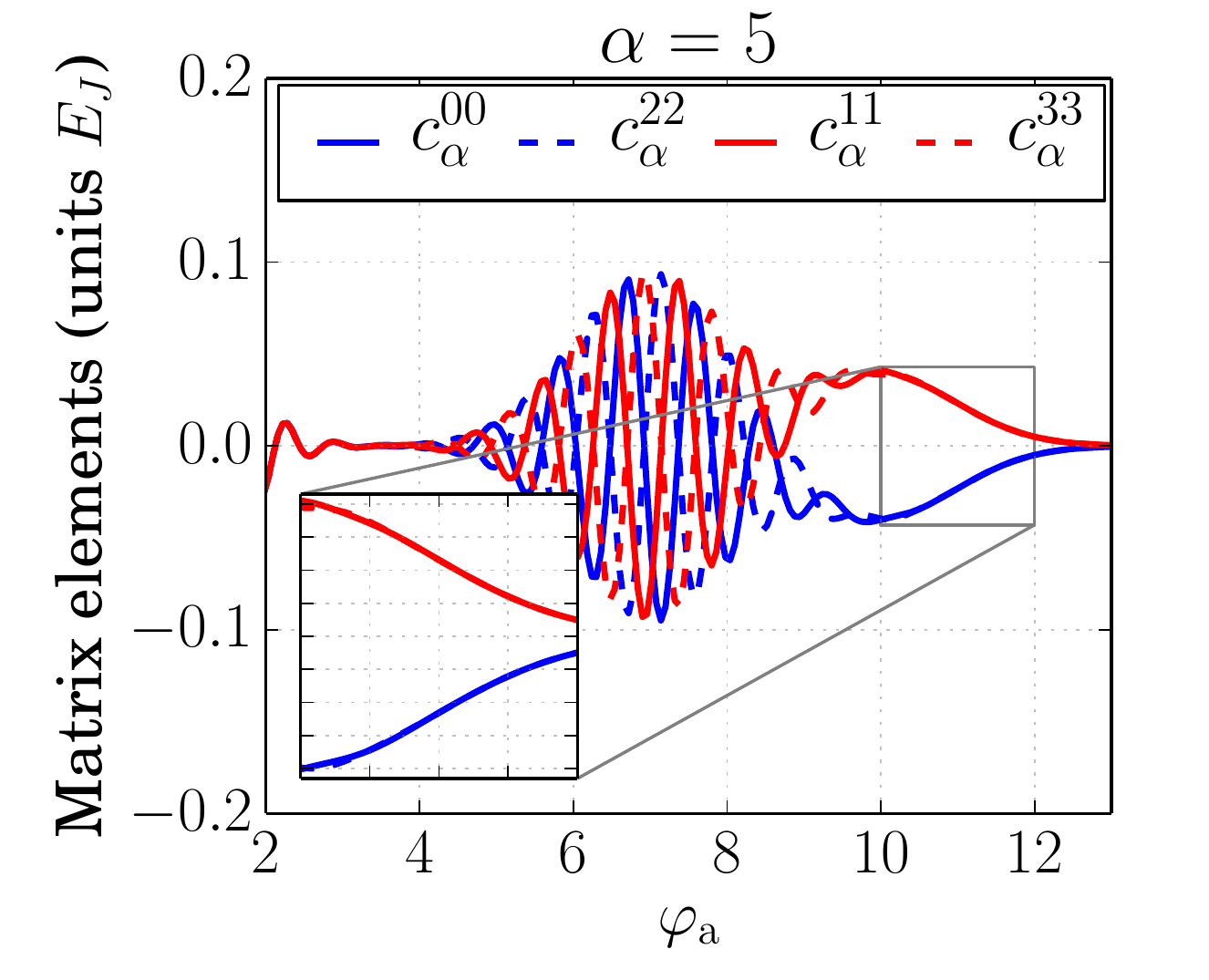}\label{fig:matrixelements4alpha5}}
\hfill
\subfloat[]{\includegraphics[width=0.37\textwidth]{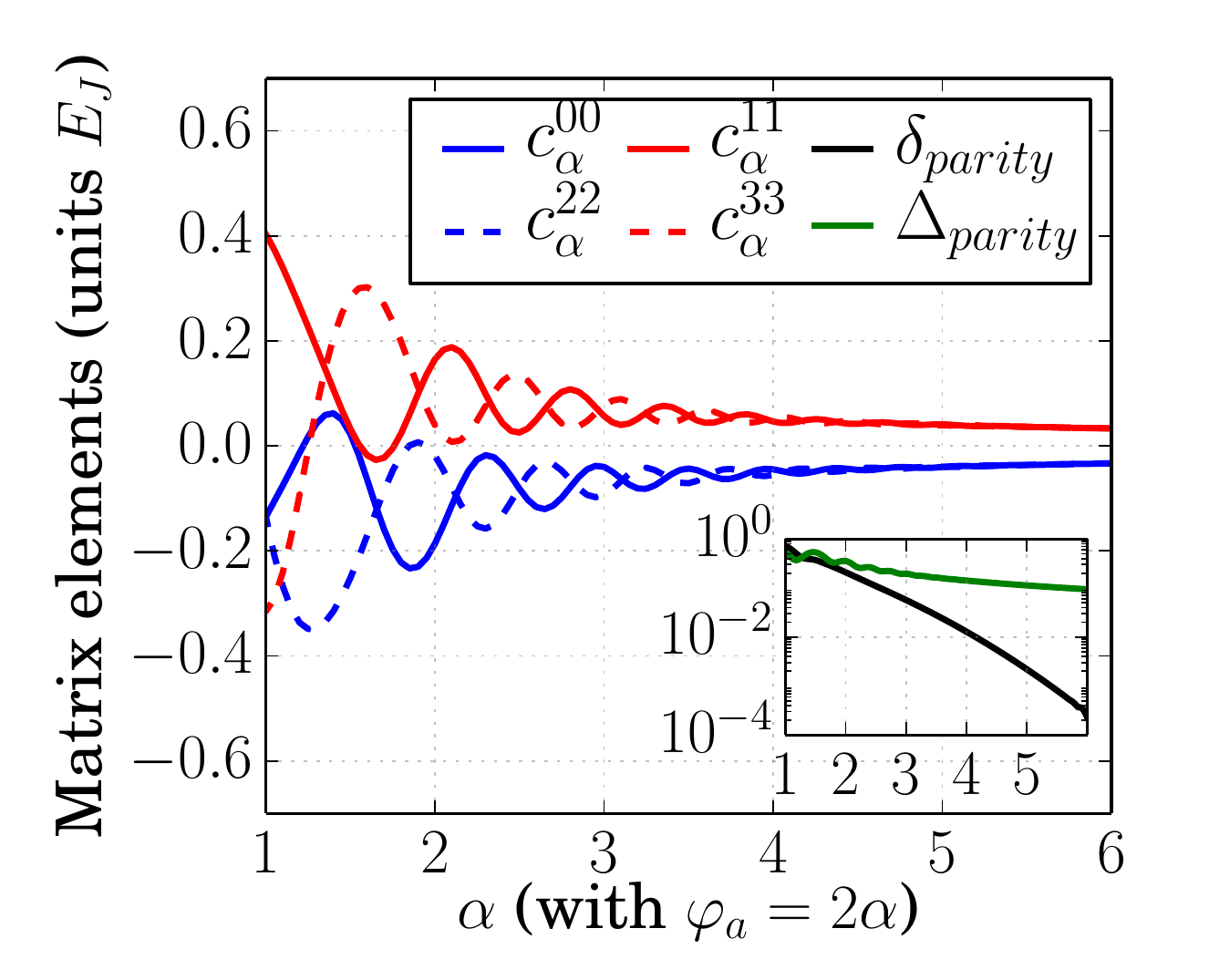}\label{fig:matrixelements4}}
  \caption{(a) Eigenvalues of $\hhh^{\text{RWA}}/E_J$ for $\varphi_a=4$. The changing signs around the Fock state $\ket{4}$ explain why under two or four-photon process, the Hamiltonian acts as a parity Hamiltonian for a coherence state $\ket{\alpha}$ with $|\alpha|=\varphi_a/2=2$. 
(b) Non-vanishing matrix elements (in units of $E_J$) of projected Hamiltonian for the two-photon driven dissipation, $c_\alpha^\pm = \bra{\CCC^{\pm}_{\alpha}} \hhh^{\text{RWA}} \ket{\CCC^{\pm}_{\alpha}}$ as a function of $\varphi_a$ ($\alpha$ being set to 2).  As shown in the inset, $c_\alpha^\pm$ take opposite  values for $3<\varphi_a<5$, indicating that the projected Hamiltonian acts as the $\sss_Z$ Pauli operator in the logical basis $\ket{\CCC_{\alpha}^{\pm}}$. 
(c) Non-vanishing matrix elements (in units of $E_J$) of projected Hamiltonian for the four-photon driven dissipation, $c_\alpha^{jj} = \bra{\CCC^{(j\text{mod}4)}_{\alpha}} \hhh^{\text{RWA}} \ket{\CCC^{(j\text{mod}4)}_{\alpha}}$ as a function of $\varphi_a$ ($\alpha$ being set to 5). We note that for $9<\varphi_a<12$ (corresponding to a window around $2\alpha$), the Hamiltonian is degenerate in each parity subspace.
%Varying $\varphi_a$, we observe three different regimes: for $\varphi_a<4$, the Hamiltonian is fully degenerate in the cat subspace and acts as the identity; for $4<\varphi_a<9$, the Hamiltonian is  non-degenerate and distinguishes the four cat states; for $9<\varphi_a<12$ which corresponds to a window around $2\alpha$, the Hamiltonian is degenerate in each parity subspace. This is the regime that we explore to ensure a parity measurement for 4-photon process. 
(d) Effect of the amplitude $|\alpha|$ on the parity-subspace degeneracy for the 4-photon process. Fixing $\varphi_a=2\alpha$, we observe that for $\alpha<4$, we deal with a non-degenerate Hamiltonian (hence the choice of $\alpha=5$ in (c)). As shown in the inset illustrates that while the parity Hamiltonian strength $\Delta_{\text{parity}}=\sqrt{(c_{\alpha}^{00}-c_{\alpha}^{11})^2+(c_{\alpha}^{22}-c_{\alpha}^{33})^2}$ decreases in $1/|\alpha|$, the parity subspace non-degeneracy $\delta_{\text{parity}}=\small{\sqrt{(c_{\alpha}^{00}-c_{\alpha}^{22})^2+(c_{\alpha}^{11}-c_{\alpha}^{33})^2}}$ decreases exponentially in $|\alpha|^2$. }
\end{figure}

In the limit $\hbar \omega_a \gg E_J$, we can apply a rotating wave approximation (RWA) to $\hhh_{\text{int}}(t)$~\cite{Gramich-Ankerhold-PRL-2013,Trif-Simon-PRB-2015,Hofer-Clerk-PRB-2016}, leading to
\begin{align} \label{eq:RWAHamiltonian}
\hhh^{\text{RWA}} = -E_J e^{-\frac{\varphi_a^2}{2}} \sum_n L_{n}(\varphi_a^2) \ket{n}\bra{n},
\end{align}
where $L_n(.)$ is the Laguerre polynomial of order $n$.
In the presence of two-photon loss, the effective Hamiltonian of the system, given by  $\hhh^{\text{RWA}}_{\MMM_{2,\alpha}}$, follows a remarkable result. Under the condition
$\varphi_a \approx 2 |\alpha|$,
the Hamiltonian takes the form of the parity Hamiltonian, i.e 
\begin{align} \label{eq:result}
\hhh^{\text{RWA}}_{\MMM_{2,\alpha}} &= -\frac{\hbar\Omega_a}{2}[ \ket{\CCC^{+}_{\alpha}} \bra{\CCC^{+}_{\alpha}}-\ket{\CCC^{-}_{\alpha}}\bra{\CCC^{-}_{\alpha}}] +\mathcal{O}(E_J e^{-\frac{\varphi_a^2}{2}}) \notag \\
&=  -\frac{\hbar\Omega_a}{2} \sss_z^L+ \mathcal{O}(E_Je^{-\frac{\varphi_a^2}{2}}),
\end{align}
where $\Omega_a$ is a function of $E_J$, $\varphi_a$ and $\alpha$. It is well approximated by
$
\Omega_a= E_J e^{-\frac{1}{2}(\varphi_a-2|\alpha|)^2}/\hbar\sqrt{\pi |\alpha| \varphi_a}
$ ~\cite{Supp}.
%Let us sketch a proof of this statement. First, at all time $t$ one can write $\bra{\CCC^{+}_{\alpha}} \hhh_{int}(t) \ket{\CCC^{+}_{\alpha}} = -\bra{\CCC^{-}_{\alpha}} \hhh_{int}(t) \ket{\CCC^{-}_{\alpha}} + \mathcal{O}(e^{-\frac{\varphi_a^2}{2}})$, from which we deduce that $\bra{\CCC^{+}_{\alpha}} \hhh^{\text{RWA}} \ket{\CCC^{+}_{\alpha}} = -\bra{\CCC^{-}_{\alpha}} \hhh^{\text{RWA}} \ket{\CCC^{-}_{\alpha}} +\mathcal{O}(e^{-\frac{\varphi_a^2}{2}})$.  In fact, the matrix elements $\bra{C^{\pm}_{\alpha}} \hhh_{int}(t) \ket{C^{\pm}_{\alpha}}$ take non-zero values only around times $t_k=(k+1/2)\pi/ \omega_a$. Note that at these times $t_k$, for $\varphi_a=2\alpha$, the Hamiltonian takes precisely the desired form, since $\hhh_{int}(t_k)=-\frac{E}{2}[\ddd(2\alpha)+\ddd(-2\alpha)]$ leads to $\hhh_{int}^{\MMM_{2,\alpha}}(t_k) = -\frac{E}{2}[ \ket{\CCC^{+}_{\alpha}} \bra{\CCC^{+}_{\alpha}}-\ket{\CCC^{-}_{\alpha}}\bra{\CCC^{-}_{\alpha}}] + \mathcal{O}(e^{-\frac{\varphi_a^2}{2}})$. Secondly, off-diagonal elements $\bra{C^{\pm}_{\alpha}} \hhh_{int}(t) \ket{C^{\mp}_{\alpha}}$ vanish with the RWA as the parity operator commutes with $\hhh^{\text{RWA}}$, which concludes the proof. In the supplemental material, we present analytical expressions for the matrix elements $c_{\alpha}^{\pm} = \bra{C^{\pm}_{\alpha}} \hhh^{\text{RWA}} \ket{C^{\pm}_{\alpha}}$ as a function of $\alpha$ and $\varphi_a$ that lead to the expression~(\ref{eq:E}) of $\Omega_a$ for $\varphi_a$ close enough to $2 |\alpha|$. 

In Fig.~\ref{fig:eigenvalues}, we plot, for $\varphi_a = 4$, the eigenvalues of $\hhh^{\text{RWA}}{/E_J}$ associated to various Fock states. Following the above {arguments}, $\hhh^{\text{RWA}}$ acts, for $\alpha \approx 2$, as a parity Hamiltonian on $\MMM_{2,\alpha}$. This can be understood through the observation of alternating signs for the eigenvalues of  $\hhh^{\text{RWA}}$ around the Fock state $\ket{4}$ corresponding to the average photon number in the coherent state $\ket{\alpha}$. Although the parity operator $\cos(\pi\aaa^\dag\aaa)$ requires also its eigenvalues to have the same module, this sign alternance is sufficient for having a parity Hamiltonian under two-photon loss. In Fig.~\ref{fig:matrixelements}, we fix $\alpha = 2$ and we plot the diagonal matrix elements $c_\alpha^\pm = \bra{\CCC^{\pm}_{\alpha}} \hhh^{\text{RWA}} \ket{\CCC^{\pm}_{\alpha}}$ as a function of $\varphi_a$ and in units of $E_J$. Note that, the off-diagonal terms $\bra{\CCC^{\pm}_{\alpha}} \hhh^{\text{RWA}} \ket{\CCC^{\mp}_{\alpha}}\equiv 0$ as $H^{\text{RWA}}$ is diagonal in the Fock states basis and does not couple even and odd manifolds. For $\varphi_a < 1.5$, $c_{\alpha}^{+}$ and $c_{\alpha}^{-}$ are roughly equal, meaning that the Hamiltonian $\hhh^{\text{RWA}}_{\MMM_{2,\alpha}}$ acts as the identity on $\MMM_{2,\alpha}$. At larger values of $\varphi_a$, $c_{\alpha}^{+}$ and $c_{\alpha}^{-}$ differ from each other. In particular, as shown in the inset of Fig.~\ref{fig:matrixelements}, around $\varphi_a = 4 = 2|\alpha|$, $c_{\alpha}^{+}$ and $c_{\alpha}^{-}$ take opposite values. In this case, $\hhh^{\text{RWA}}_{\MMM_{2,\alpha}}$ becomes proportional to $\sss_z^L$.

From the construction of a single-mode parity Hamiltonian, acting as a $\sss_z^L$ Pauli operator in the logical basis,  stems an immediate route to build a joint-parity Hamiltonian of two cavity modes $\aaa$ and $\bbb$ both subject to two-photon dissipation. Considering two cavity modes $\aaa$ and $\bbb$ coupled to a Josephson junction, the interaction Hamiltonian reads $\hhh_{int}(t) = -E_J \cos[\varphi_a(\aaa e^{-i \omega_a t }+ c.c)+\varphi_b(\bbb e^{-i \omega_b t }+ c.c)]$. The mode frequencies $\omega_a$ and $\omega_b$ are off-resonant so that we can apply the RWA (more precisely, one needs to choose these frequencies in a way to avoid also high-order resonances)
\begin{align}
%\label{eq:twomodeRWA}
\hhh^{\text{RWA}} = & -E_J e^{-\frac{\varphi_a^2+\varphi_b^2}{2}} \notag  \\ 
& \sum_{n_a,n_b} L_{n_a}(\varphi_a^2) L_{n_b}(\varphi_b^2) \ket{n_a, n_b}\bra{n_a, n_b}.\notag
\end{align}
Similarly to the single-mode case, if both $\aaa$ and $\bbb$ are  high-impedance modes and are subject to two-photon loss, one can choose $|\alpha|\approx \varphi_a/2$ and $|\beta|\approx \varphi_b/2$, such that the confined Hamiltonian takes the form
$
\hhh^{\text{RWA}}_{\MMM_{2,\alpha,\beta}} =  -\frac{\hbar\Omega_{a,b}}{2}\sss_Z^{a,L}\otimes\sss_Z^{b,L}, 
$
where $\Omega_{a,b} = \hbar\Omega_a \Omega_b /2E_J$, $\sss_Z^{a(b),L}=\ket{\CCC_{\alpha(\beta)}^+}\bra{\CCC_{\alpha(\beta)}^+}-\ket{\CCC_{\alpha(\beta)}^-}\bra{\CCC_{\alpha(\beta)}^-}$.

We have seen that under two-photon loss, the Hamiltonian $\hhh^{\text{RWA}}$ acts as a parity Hamiltonian. Remarkably, this result also holds in the presence of four-photon loss, where the dynamics is confined to the larger manifold $\MMM_{4,\alpha} = \MMM_{2,\alpha} \oplus \MMM_{2,i\alpha}$. More precisely, for $\varphi_a\approx 2|\alpha|$, the projection of $\hhh^{\text{RWA}}$ on $\MMM_{4,\alpha}$ satisfies ${\hhh^{\text{RWA}}_{\MMM_{4,\alpha}} =-\hbar\Omega_a/2 ~(\Pi_{\MMM_{4,\alpha}}\cos(\pi\aaa^\dag\aaa)\Pi_{\MMM_{4,\alpha}} + O(e^{-\xi |\alpha|^2}))}$, with $\xi = (\sqrt 2-1)^2\approx 0.17$~\cite{Supp}. The undesired term that scales as $e^{-\xi |\alpha|^2}$  lifts the degeneracy within the parity subspaces. This non-degeneracy is however suppressed exponentially with cat size $|\alpha|^2$, while the effective Hamiltonian strength decreases only linearly in $|\alpha|^{-1}$. Therefore for large enough $\alpha$'s we still achieve an effective parity Hamiltonian. This is illustrated in Fig.~\ref{fig:matrixelements4}, where we plot the diagonal matrix elements $c_{\alpha}^{jj}=\bra{\CCC_\alpha^{j\text{mod}4}}\hhh^{\text{RWA}}\ket{\CCC_\alpha^{j\text{mod}4}}$. As shown in the inset,  the non-degeneracy vanishes rapidly and around values of $|\alpha|=4$, $\hhh^{\text{RWA}}_{\MMM_{4,\alpha}}$ is well-approximated by a parity Hamiltonian. The perfect degeneracy, for cat states of smaller amplitude, can also be achieved by introducing more junctions providing more degrees of freedom (see~\cite{Supp} for more details). Similarly to Fig.~\ref{fig:matrixelements}, in Fig.~\ref{fig:matrixelements4alpha5}, we fix $\alpha = 5$ (for which, as analyzed in Fig.~\ref{fig:matrixelements4}, the parity subspace degeneracy is ensured) and we plot the diagonal matrix elements $c_\alpha^{jj}$ as a function of $\varphi_a$ and in units of $E_J$. As shown in the inset, around $\varphi_a = 10 = 2|\alpha|$, we achieve an effective  Hamiltonian of the form $\boldsymbol{\pi}_{4\text{ph}}=\ket{\CCC_\alpha^{(0\text{mod}4)}}\bra{\CCC_\alpha^{(0\text{mod}4)}}+\ket{\CCC_\alpha^{(2\text{mod}4)}}\bra{\CCC_\alpha^{(2\text{mod}4)}}-\ket{\CCC_\alpha^{(1\text{mod}4)}}\bra{\CCC_\alpha^{(1\text{mod}4)}}-\ket{\CCC_\alpha^{(3\text{mod}4)}}\bra{\CCC_\alpha^{(3\text{mod}4)}}$.

\begin{figure}[!tbp]
  \centering
\subfloat[]{\includegraphics[width=0.36\textwidth]{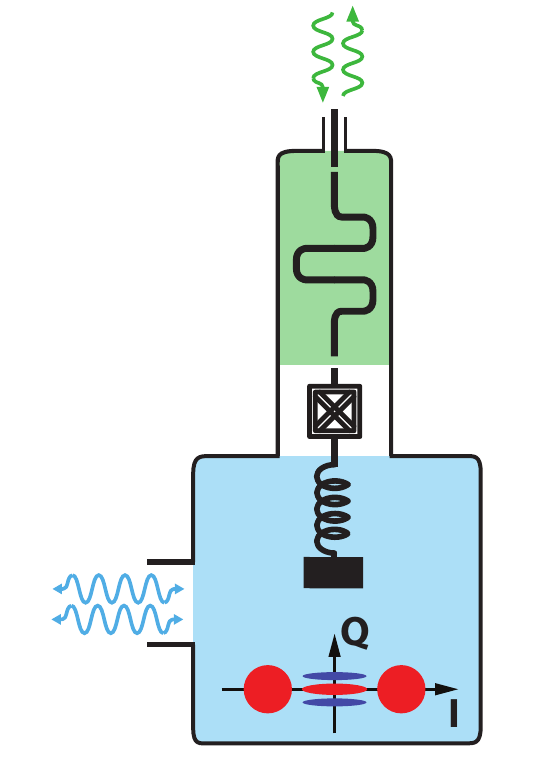}\label{fig:onemodesetup}}
\hfill
\subfloat[]{\includegraphics[width=0.27\textwidth]{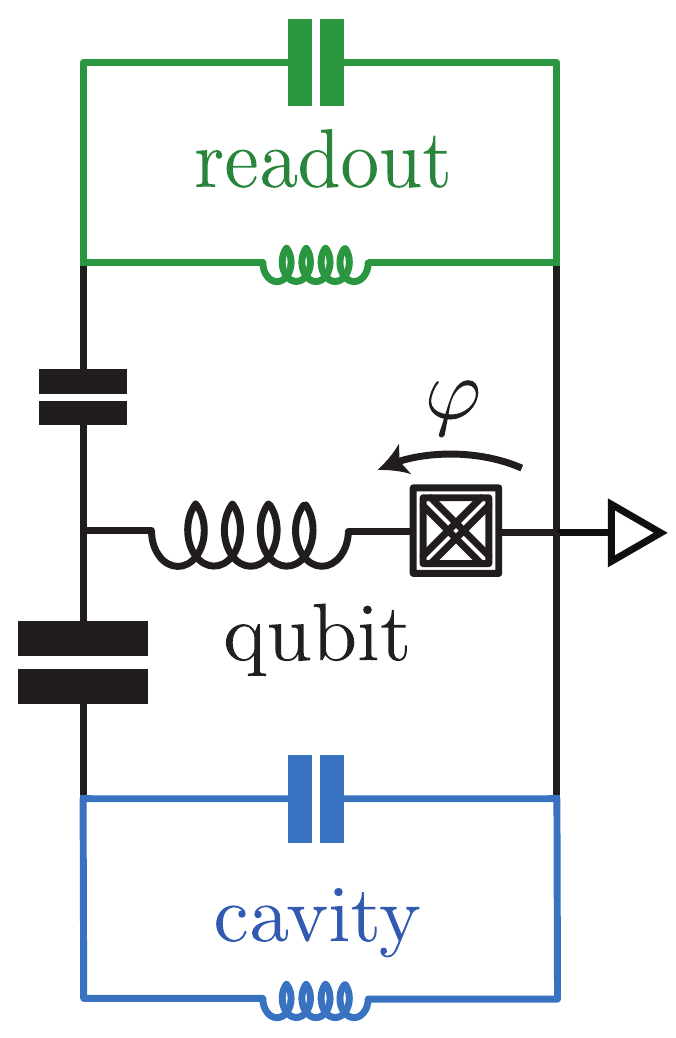}\label{fig:circuit}}
\hfill
\subfloat[]{\includegraphics[width=0.37\textwidth]{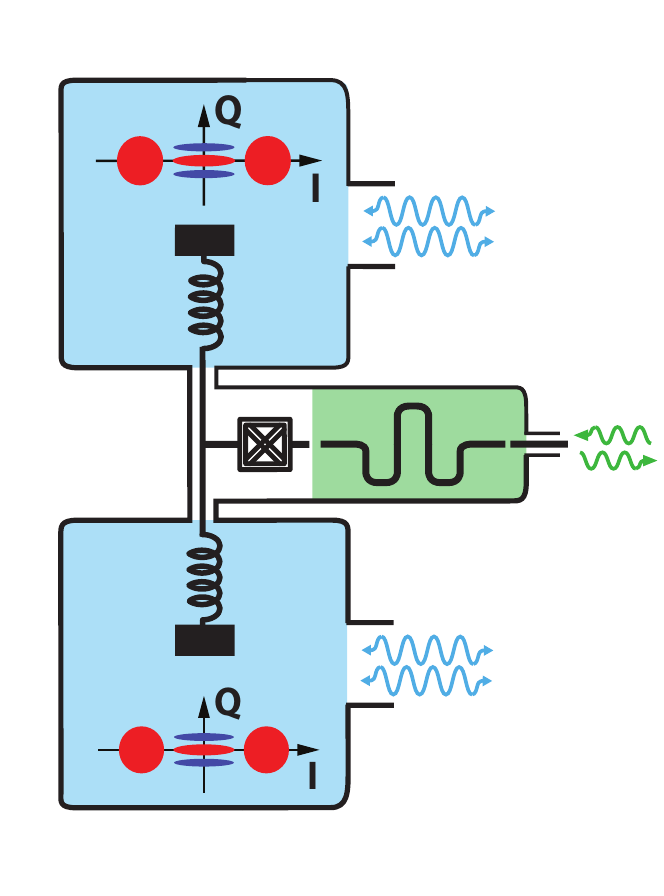}\label{fig:twomodesetup}}
  \caption{(a) Schematic of a realization of a single-mode continuous parity measurement in presence of two-photon driven dissipation. Similarly to~\cite{Touzard-et-al-APS-2016}, on one side we mediate a two-photon dissipation of the storage high-Q cavity mode and on the other, we couple to a low-Q readout mode through a high-impedance Josephson circuit. 
 (b) Electrical circuit equivalent, without the two-photon driven dissipation. The cavity (blue) and readout (green) modes are modeled by LC oscillators, and are capacitively coupled to a high impedance Josephson circuit mode. This Josephson mode consists of a large superinductance, formed from an array of large Josephson junctions (as in the fluxonium), in series with a nonlinear circuit element, depicted as a cross-hatched box. 
 (c) Schematic design of the joint-parity measurements between two high-Q cavity modes under two-photon driven dissipation, inspired from~\cite{Wang-Gao-Science-2016} and based on an extension of (a).}
\end{figure}

Following the same idea as in the usual dispersive measurements of superconducting qubits~\cite{Wallraff-Nature-2004}, one can perform a continuous quantum non-demolition measurement of the above observables, $\sss_Z^L$ and $\sss_Z^L\otimes\sss_Z^L$ for the two-photon dissipation scheme, and $\boldsymbol{\pi}_{4\text{ph}}$ for the four-photon dissipation. This can be done by coupling an extra off-resonant readout mode to  the same junction. This mode is then driven at its resonance (see Figs.~\ref{fig:onemodesetup}-c) and the measurement outcome is imprinted on the phase or/and the amplitude of the reflected signal. More precisely, by coupling a driven readout mode $\ccc$ to the junction, and in the case of $\varphi_c \sqrt{n_c}\ll 1$ (here $n_c$ denotes the average number of readout photons and this requirement is equivalent to assuming $n_c\ll n_{\text{crit}}$, the critical number for dispersive approximation~\cite{Gambetta-et-al-08}), we achieve the following effective Hamiltonians:
\begin{align} \label{eq:measurementHamiltonian}
\hhh^{\text{disp}}_{\MMM_{2,\alpha}} & \approx -\frac{\hbar\widetilde\Omega_a}{2}\sss_Z^a +\frac{\hbar\chi_a}{2} \sss_Z^a \ccc^\dag\ccc+\hhh_{\text{drive}}(t), \notag\\
\hhh^{\text{disp}}_{\MMM_{2,\alpha,\beta}} & \approx -\frac{\hbar\widetilde\Omega_{a,b}}{2}\sss_Z^a\sss_Z^b +\frac{\hbar\chi_{a,b}}{2} \sss_Z^a \sss_Z^b \ccc^\dag\ccc+\hhh_{\text{drive}}(t), \notag\\
\hhh^{\text{disp}}_{\MMM_{4,\alpha}} & \approx -\frac{\hbar\widetilde\Omega_a}{2}\boldsymbol{\pi}_{4\text{ph}} +\frac{\hbar\chi_a}{2} \boldsymbol{\pi}_{4\text{ph}} \ccc^\dag\ccc+\hhh_{\text{drive}}(t).
\end{align}
Here $\hhh_{\text{drive}}(t)=\hbar(\epsilon_c(t)\ccc^\dag+\epsilon_c^*(t)\ccc)$, $\widetilde\Omega_a=e^{-\varphi_c^2/2}\Omega_a$, $\widetilde\Omega_{a,b}=e^{-\varphi_c^2/2}\Omega_{a,b}$, $\chi_a=\widetilde\Omega_a \varphi_c^2$, $\chi_{a,b}=\widetilde\Omega_{a,b} \varphi_c^2$. 

\begin{figure}[!tbp]
\centering
{\includegraphics[width=.7\textwidth]{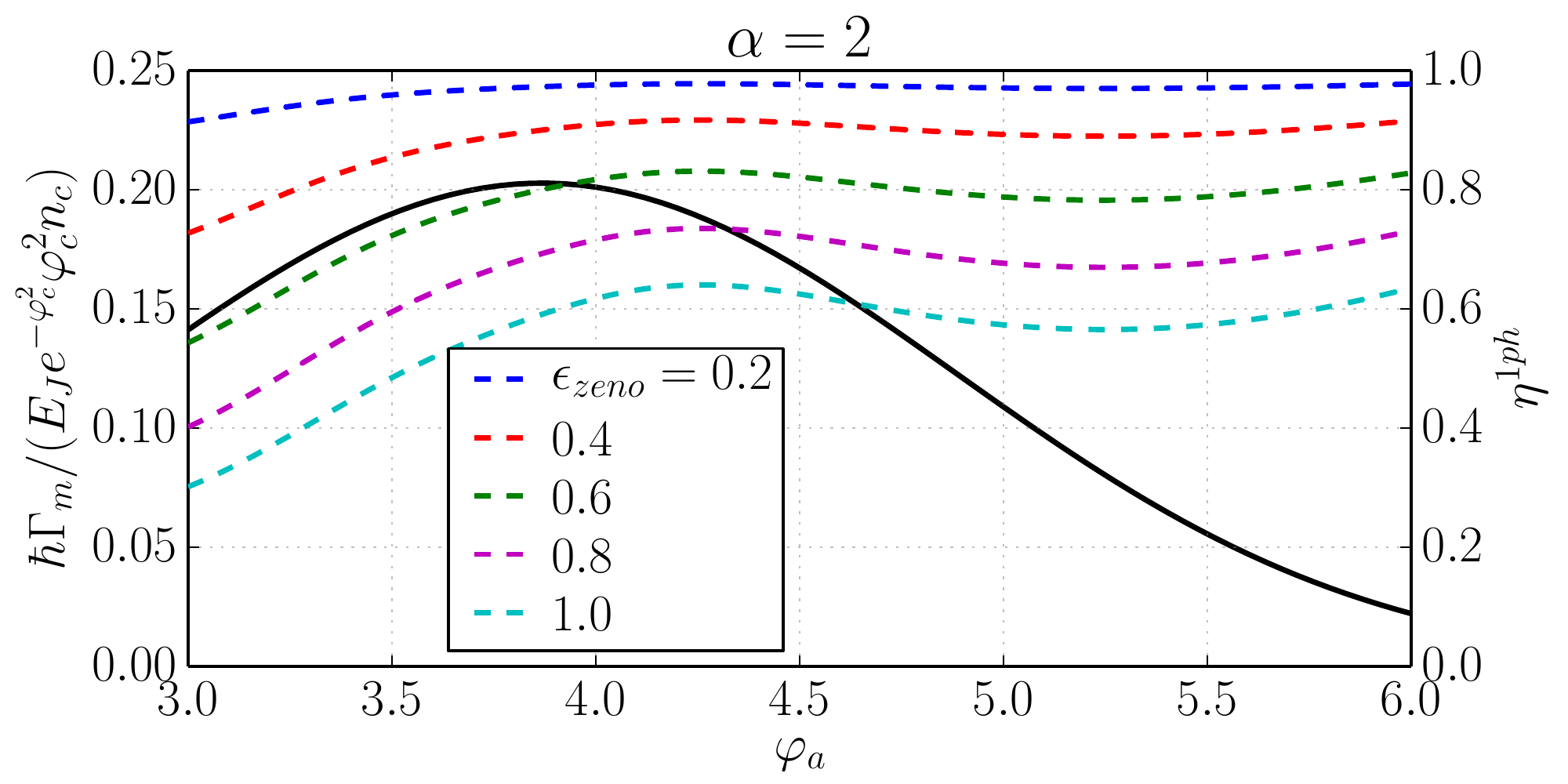}}
\caption{The left axis corresponding to the black straight curve illustrates the measurement rate in the single-mode case~\eqref{eq:Gamma} in units of $E_J/\hbar$ and renormalized by the parameters of the readout mode. Fixing $\alpha=2$ and varying $\varphi_a$, we observe an optimal measurement rate around $\varphi_a=2\alpha$. The right axis, corresponding to the colored dashed curves, illustrate the efficiency of the measurement limited by the higher order Zeno effects (simulating~\eqref{eq:master}). Here, we fix $\varphi_c=.1$ and the number of readout photons $n_c=1$, and we plot $\eta_{1\text{ph}}=\Gamma_m^{1\text{ph}}/(\Gamma_m^{1\text{ph}}+\Gamma_Z)$. Varying $\varphi_a$, we observe that this efficiency achieves a local optimum near $\varphi_a=2\alpha $ corresponding also to the optimum measurement rate $\Gamma_m$. Also, by decreasing the Zeno parameter $\epsilon_{\text{zeno}}=E_J/{\hbar\kappa_{2ph}}$, this higher-order effect can be suppressed.  \label{fig:QNDness}}
\end{figure}

The first terms in the above Hamiltonians simply induce deterministic rotations in the associated parity subspaces, whereas the second terms correspond to frequency pulls on mode $\ccc$ that depend on the values of associated observables. By driving the mode $\ccc$ at resonance, so that $\hhh_{\text{drive}}$ is time-independent, the measurement outcome is imprinted on the phase of the pointer coherent state.  Taking $\kappa_c$ to be the dissipation rate of $\ccc$ induced by its coupling to a readout transmission line, the measurement rate is optimal when $\kappa_c=\chi_{a}$ ($\chi_{a,b}$ for joint-parity measurement)~\cite{Gambetta-et-al-06}.  This optimal rate is given by (see Fig.~\ref{fig:QNDness})
\begin{align} \label{eq:Gamma}
\Gamma_m^a &= \bar{n}_c\chi_a= \bar{n}_c \varphi_c^2 e^{-\frac{\varphi_c^2}{2}} \frac{E_J}{\hbar} \frac{e^{-\frac{1}{2}(\varphi_a-2|\alpha|)^2}}{\sqrt{\pi |\alpha| \varphi_a}} ,\\ 
\Gamma_m^{a,b} &= \bar{n}_c\chi_{a,b}= \bar{n}_c \varphi_c^2 e^{-\frac{\varphi_c^2}{2}} \frac{E_J}{\hbar} \frac{e^{-\frac{1}{2}(\varphi_a-2|\alpha|)^2-\frac{1}{2}(\varphi_b-2|\beta|)^2}}{2\pi \sqrt{|\alpha\beta| \varphi_a\varphi_b}}.\notag
\end{align}
Practical realization of a high impedance cavity mode, satisfying $\varphi_a \approx 2 |\alpha|$, poses a notable challenge. To see this, note that this relation for $\alpha = 2$ requires an impedance $Z_a = 32 R_Q$. For comparison, typical superconducting cavities have impedances $0.1 R_Q < Z_a < R_Q$ \cite{Paik-et-al-PRL2011, Reagor2016}. However, much larger impedances $Z \sim 8 R_Q$ have been produced using devices comprising superinductances (fabricated from arrays of large Josephson junctions), such as in the fluxonium qubit \cite{Masluk-Devoret-PRL-2012, Pop-Devoret-Nature-2014}.

In our proposed experimental system (see Figs.~\ref{fig:onemodesetup}-b, a fluxonium-based qubit mode composed of a superinductance in series with a nonlinear circuit element is capacitively coupled to two cavities. This nonlinear circuit element is assumed to have a Josephson junction-like Hamiltonian of the form 
\begin{equation*}
\mathbf{H}_\text{el} = 4 E_C \mathbf{n}^2 - E_J \cos \mu \boldsymbol \varphi,
\end{equation*}
where $\mathbf{n}$ is the number of Cooper pairs across the element, $\boldsymbol\varphi$ is the superconducting phase, $E_C$ is the charging energy, $E_J$ is the Josephson energy, and $\mu$ is an integer-valued parameter determined by the implementation. It may be worthwhile to realize $\mu > 1$, and this could be achieved by circuits similar to those proposed in~\cite{Ioffe2002,Brooks2013}. This transforms the effective cavity impedance according to $Z_a \rightarrow \mu^2 Z_a$, making the relation $\varphi_a \approx 2 |\alpha|$ much easier to satisfy. The details of this strategy will be described in a forthcoming publication.

Let us now study the limitations of such a measurement protocol. Here we have made a few approximations and the main limitations are due to second order effects. The first one concerns the RWA. Indeed, dealing with high-impedance modes one needs to be  cautious about  higher order resonances. While in the single-mode case, such second-order effects lead to a slight modification of the measurement rate, in the two-mode case, they could lead to a small dephasing within the parity subspaces (see~\cite{Supp}).  These effects could be minimized by a careful choice of resonance frequencies. 

Another limitation concerns the Zeno approximation. We have considered that under the two-photon process, the confined dynamics is given by the projected Hamiltonian $\hhh^{\text{RWA}}_{\MMM_{2,\alpha}}$. This corresponds to a first order Zeno dynamics approximation in $\epsilon_{\text{zeno}} = E_J/{\hbar\kappa_{2ph}}$~\cite{Azouit-CDC-2015,Azouit-CDC-2016}. The second order correction in $\epsilon_{\text{zeno}}$ induces a dephasing in the basis $\{\ket{\CCC_\alpha^\pm}\}$ occurring at a rate $\Gamma_Z = r(\alpha,\varphi_a) \epsilon_{\text{zeno}}^2 {\kappa_{2ph}}$, where the numerical factor $r(\alpha,\varphi_a)$ can be derived from~\cite{Azouit-CDC-2016}.  This could be seen as an inefficiency in the measurement, where a constant part (independent of the number of readout photons) of the measurement signal is lost through the two-photon decay channel. Here, we analyze numerically this second order effect by simulating the master equation
\begin{equation}\label{eq:master}
\frac{d\rrr}{dt} = -i[\hhh^{\text{RWA}},\rrr]+\kappa_{2ph}\ddd[\aaa^2-\alpha^2](\rho) 
\end{equation}
where the single mode Hamiltonian, $\hhh^{\text{RWA}}$, is given by~(\ref{eq:RWAHamiltonian}) and $\rrr(0)=\frac{1}{2}(\ket{\CCC_\alpha^+}+\ket{\CCC_\alpha^-})(\bra{\CCC_\alpha^+}+\bra{\CCC_\alpha^-})$. 
Taking $\alpha=2$, and varying $\varphi_a$ and the Zeno parameter $\epsilon_{\text{zeno}}$, we look at the decay of purity with time. This corresponds to a dephasing due to higher order Zeno dynamics well-approximated by $\Gamma_Z$. We illustrate in Fig.~\ref{fig:QNDness} (dashed lines corresponding to the right axis) the value of $\eta_{1\text{ph}}=\Gamma_m^{1\text{ph}}/(\Gamma_m^{1\text{ph}}+\Gamma_Z)$ corresponding to the measurement efficiency when the number of readout photons is fixed to $n_c=1$. Indeed, as $\Gamma_Z$ does not depend on the number of readout photons while $\Gamma_m$ increases linearly in $n_c$, this efficiency improves for higher number of readout photons.  We observe that for a given Zeno parameter $\epsilon_{\text{zeno}}$, the point $\varphi_a\approx 2\alpha$ corresponding to the optimal measurement rate $\Gamma_m$, is also a local optimum of the efficiency. We do not account for the readout mode $\ccc$ in these simulations, as it does not contribute to higher-order Zeno approximations and therefore to $\Gamma_Z$. Note that, this measurement inefficiency is the only detrimental effect of such higher-order dynamics. As the Hamiltonian $\hhh^{\text{RWA}}$ is diagonal in the Fock states basis, it does not change the parity and therefore do not lead to any bit-flip type error of the logical qubit. 

We can perform a similar analysis for the two-mode joint-parity measurement protocol. While higher order Zeno dynamics cannot lead to any change of photon number parities (Hamiltonian is diagonal in the Fock states basis), in principle, it can lead to a dephasing for each logical qubit. However, to a very good approximation (exponentially precise in $|\alpha|^2$), such a dephasing occurs in a correlated manner, giving rise to a dissipation channel of the form $\sss_Z^a\otimes\sss_Z^b$ (see~\cite{Supp}). This means that such higher order effects do not induce any decoherence within a given joint-parity subspace. Therefore starting from  $c_{++}\ket{\CCC_\alpha^+,\CCC_\alpha^+}+ c_{--}\ket{\CCC_\alpha^-,\CCC_\alpha^-}+ c_{+-}\ket{\CCC_\alpha^+,\CCC_\alpha^-}+ c_{-+}\ket{\CCC_\alpha^-,\CCC_\alpha^+}$, the measurement will project the state on one of the two parity states $c_{++}\ket{\CCC_\alpha^+,\CCC_\alpha^+}+ c_{--}\ket{\CCC_\alpha^-,\CCC_\alpha^-}$ or $c_{+-}\ket{\CCC_\alpha^+,\CCC_\alpha^-}+ c_{-+}\ket{\CCC_\alpha^-,\CCC_\alpha^+}$ without affecting the purity of these states. We thus deal with a quantum non-demolition measurement (with non-unit efficiency) of joint parity. 

We have shown how to achieve continuous quantum non-demolition measurement of three parity-type observables for harmonic oscillators. We focus on the case of multi-photon driven dissipative systems previously introduced for universal quantum computation with cat-qubits~\cite{Mirrahimi-al-NJP-2014}. The three observables consist of $\sss_Z^a=\ket{\CCC_\alpha^+}\bra{\CCC_\alpha^+}-\ket{\CCC_\alpha^-}\bra{\CCC_\alpha^-}$ for a single-mode under two-photon process, joint-parity $\sss_Z^a\otimes\sss_Z^b$ for two modes under two-photon process, and $\boldsymbol{\pi}_{4\text{ph}}=\ket{\CCC_\alpha^{(0\text{mod}4)}}\bra{\CCC_\alpha^{(0\text{mod}4)}}+\ket{\CCC_\alpha^{(2\text{mod}4)}}\bra{\CCC_\alpha^{(2\text{mod}4)}}-\ket{\CCC_\alpha^{(1\text{mod}4)}}\bra{\CCC_\alpha^{(1\text{mod}4)}}-\ket{\CCC_\alpha^{(3\text{mod}4)}}\bra{\CCC_\alpha^{(3\text{mod}4)}}$ under four-photon process. The continuous and QND measurement of these observables play a central role towards scalable fault-tolerant architectures for universal quantum computation. 
We also propose a possible implementation of these measurements through the high-impedance coupling of the cavity mode(s) to a Josephson junction. 
While the focus of this Letter is on Zeno dynamics induced by multi-photon driven dissipation, the scheme could also be adapted to non-dissipative cases such as~\cite{Puri-Blais-2016}. Indeed, in presence of strong Kerr type non-linearities, the Hamiltonian perturbation due to high-impedance coupling to a Josephson circuit results in the creation of a parity Hamiltonian. More precisely, considering a cavity subjected to strong self-Kerr effect and a two-photon drive, the Hamiltonian, in the interaction picture, is given by $\hhh_0 = -\hbar K({\aaa^\dag}^2-\frac{\EEE_p^*}{K})({\aaa}^2-\frac{\EEE_p}{K})$, with $K$ the self-Kerr coefficient, and $\EEE_p$ the two-photon drive strength~\cite{Puri-Blais-2016}. The 2D-manifold $\MMM_{2,\alpha} = \text{span} \{ \ket{\CCC_\alpha^\pm} \}$ is a doubly degenerate eigenspace of $\hhh_0$, separated from the other eigenspaces by an energy gap of order $4\hbar |\EEE_p|$. Considering the first order effect of a perturbative Hamiltonian $\hhh_1=\hhh^\text{RWA}$ (see expression (\ref{eq:RWAHamiltonian})) with $|| \hhh_1|| \ll 4\hbar  |\EEE_p|$, we lift the degeneracy of $\MMM_{2,\alpha}$, leading to two non-degenerate eigenstates approximately given by $\ket{\CCC_\alpha^+}$ and $\ket{\CCC_\alpha^-}$. This implies that we have achieved an effective $\sss_z$ Hamiltonian on the logical basis of cat states.

This research was supported by  Inria's DPEI under the TAQUILLA associated team and by ARO under Grant No. W911NF-14-1-0011.

\newpage
\begin{center}
\large{\bf Supplemental Material}% Force line breaks with \\
\end{center}

\section{The Rotating-Wave Approximation (RWA) : Derivation and validity}

\subsection{Derivation of the hamiltonian ${\hhh^{\text{RWA}}}$ : one-mode case and two-mode case}
\label{sec:derivationRWA}

In the case of a single cavity mode coupled to a Josephson junction, the hamiltonian in the interaction picture reads $\hhh_\text{int}(t) = -E_J \cos(\varphi_a(\aaa e^{-i\omega_a t}+\aaa^\dag e^{i \omega_a t}))=-E_J/2(\DDD[c_a(t)]+\DDD[-c_a(t)])$, where $c_a(t) = i \varphi_a e^{i\omega_a t}$. We can expand the displacement operator, 
\begin{align} \label{eq:onemodeHintsupp}
\DDD[c_a(t)] = \sum \limits_{l_a=0}^\infty \amaj(l_a)(-\aaa e^{-i\omega_a t })^{l_a}+ \sum \limits_{l_a=1}^\infty (\aaa^\dag e^{i\omega_a t })^{l_a} \amaj(l_a),
\end{align}
where $\amaj (l_a) = \varphi_a^{l_a} e^{-\frac{\varphi_a^2}{2}} \sum \limits_{n_a=0} \frac{n_a!}{(n_a+l_a)!} L_n^{(l_a)}(\varphi_a^2) \ket{n_a}\bra{n_a} $ is a hermitian operator~\cite{Hofer-Clerk-PRB-2016-sup}. Here, $L_n^{(l_a)}$ is the generalized Laguerre polynomial of order $n$ and parameter $l_a$. The first order RWA, given by the formula $\hhh^{\text{RWA,1}} = \overline{\hhh(t)}$ where $\overline{\hhh(t)} = \lim \limits_{T \rightarrow \infty} \frac{1}{T} \int \limits_0^T \hhh(t) dt $, reads in our case, 
\begin{align} \label{eq:onemodeRWAsupp}
\hhh^{\text{RWA,1}} = -E_J e^{-\frac{\varphi_a^2}{2}} \sum_n L_{n}(\varphi_a^2) \ket{n}\bra{n}.
\end{align}

In the two-mode case, the hamiltonian reads in the interaction picture $\hhh_\text{int}(t) = -E_J/2(\DDD[c_a(t)]\DDD[c_b(t)]+h.c.)$, where $c_a(t) = i \varphi_a e^{i\omega_a t}$ and $c_b(t) = i \varphi_b e^{i\omega_b t}$, i.e 
\begin{multline} \label{eq:twomodeHintsupp}
\hhh_\text{int}(t)= -\frac{E_J}{2}\big[  \big( \sum \limits_{l_a=0}^\infty \amaj(l_a)(-\aaa e^{-i\omega_a t })^{l_a}+ \sum \limits_{l_a=1}^\infty (\aaa^\dag e^{i\omega_a t })^{l_a} \amaj(l_a) \big)\\ \big(  \sum \limits_{l_b=0}^\infty \bmaj(l_b)(-\bbb e^{-i\omega_b t })^{l_b}+ \sum \limits_{l_b=1}^\infty (\bbb^\dag e^{i\omega_b t })^{l_b} \bmaj(l_b) \big)  
+ h.c. \big]
\end{multline}
Similarly to $\amaj (l_a)$, we have defined $\bmaj (l_b) = \varphi_b^{l_b} e^{-\frac{\varphi_b^2}{2}} \sum \limits_{n_b=0} \frac{n_b!}{(n_b+l_b)!} L_n^{(l_b)}(\varphi_b^2) \ket{n_b}\bra{n_b}$. The frequencies $\omega_a$ and $\omega_b$ are taken to be incommensurate, meaning that $l_a \omega_a \neq l_b \omega_b$ for all $l_a,~l_b >0$. In practice, we require that the modes are sufficiently off resonant to avoid high order photon exchange terms. Under this assumption, the only non-rotating term corresponds to $l_a=0$ and $l_b=0$, which leads to
\begin{equation}
\label{eq:twomodeRWAsupp}
\hhh^{\text{RWA,1}} =  -E_J e^{-\frac{\varphi_a^2+\varphi_b^2}{2}} \sum_{n_a,n_b} L_{n_a}(\varphi_a^2) L_{n_b}(\varphi_b^2) \ket{n_a, n_b}\bra{n_a, n_b}.
\end{equation}
As mentioned in the main text, some of the high order terms eventually approach resonances. This effect, which is accounted for in the second order RWA, is studied in section~\ref{sec:validityRWA}.

%\begin{figure}[h]
%  \centering
%    \centering
%  \subfloat[]{\includegraphics[width=0.5\textwidth]{Figures/rabioscillationonemode.pdf}\label{fig:rabioscillationonemode}}
%  \hfill
%  \subfloat[]{\includegraphics[width=0.5\textwidth]{Figures/rabioscillationfourphoton.pdf}\label{fig:fourphotonlossoscillation}}
%    \hfill
%  \caption{(a) :  (b) : .}
%\end{figure}

\subsection{Validity of the RWA}
\label{sec:validityRWA}

As we consider high impedance modes, we need to be careful about the validity of the RWA. Indeed, non resonant high order terms in the development of the cosine (see expressions~(\ref{eq:onemodeHintsupp}) and~(\ref{eq:twomodeHintsupp})) have larger amplitudes than in the case of low impedance modes, and can affect the dynamics throughout higher order interaction terms. The effect of these terms can be evaluated though the second order correction of the RWA, resulting in the hamiltonian $\hhh^{\text{RWA,2}}$, given by the following expression~\cite{mirrahimi-notes,sanders-verhulst-book}
\begin{align} \label{eq:RWA2formulasupp}
\hhh^{\text{RWA,2}}=\hhh^{\text{RWA,1}}-i \overline{(\hhh(t)-\hhh^{\text{RWA,1}}) \int \limits_0^t du (\hhh(u)-\hhh^{\text{RWA,1}})}.
\end{align}  
The hamiltonian remains diagonal in the Fock states basis at the second order, so that the second order correction induces at most a shift of the hamiltonian eigenvalues. In the two-photon process for the single-mode case, this merely leads to a small modification of the measurement rate $\Gamma_m$. Moreover, as the smallest frequency of non-secular terms is given by $\omega_a \gg E_J$ (see expression~(\ref{eq:onemodeHintsupp})), this shift is small relatively to the eigenvalues. 

In the case of two modes $\aaa$ and $\bbb$ with incommensurate frequencies $\omega_a$ and $\omega_b$, the second order hamiltonian $\hhh^{\text{RWA,2}}$ is also diagonal in the Fock states basis. As a direct consequence, its projection ${\hhh^{\text{RWA,2}}_{\MMM_{2,\alpha,\beta}}}$ on the manifold ${\MMM_{2,\alpha,\beta}}$ is a linear combination of the operators $\III_a \otimes \III_b,  \sss_Z^a \otimes \III_b,  \III_a \otimes \sss_Z^b$ and $\sss_Z^a \otimes \sss_Z^b$, where $\III_{a(b)}$ is defined as the identity on ${\MMM_{2,\alpha(\beta)}}$ . Unlike in the single-mode case where this energy shift merely modifies the measurement rate, here it can also lead to an unwanted dephasing within the parity subspaces. Indeed, while for the first order RWA, ${\hhh^{\text{RWA,1}}_{\MMM_{2,\alpha,\beta}}} \propto  \sss_Z^a \otimes \sss_Z^b$, the projection ${\hhh^{\text{RWA,2}}_{\MMM_{2,\alpha,\beta}}}$ of the second order hamiltonian $\hhh^{\text{RWA,2}}$ on the manifold ${\MMM_{2,\alpha,\beta}}$ can acquire non-zero components on $\sss_Z^a \otimes \III_b$ and $ \III_a \otimes \sss_Z^b$. The measurement would therefore lead to a dephasing in two-qubit parity subspaces. Besides, the second order energy shifts for this second order approximation can, in principle, be large with respect to single-mode case. This is due to the fact that some high order terms in $l_a$ and $l_b$ (see expression~(\ref{eq:twomodeHintsupp})) become close to resonance. By inserting the expression of $\hhh(t)$ given in eq.~(\ref{eq:twomodeHintsupp}) into eq.~(\ref{eq:RWA2formulasupp}), we derive $\hhh^{\text{RWA,2}}$
\begin{align*} 
\hhh^{\text{RWA,2}} =& -E_J \hhh^{\text{RWA,1}}  \\
&+ E_J^2 \sum \limits_{l_a,l_b \geq 0,~(l_a,l_b)\neq(0,0)} \frac{(1+(-1)^{l_a+l_b})}{2(l_a \omega_a+l_b \omega_b)}[({\aaa^\dag}^{l_a} \amaj (l_a)^2 \aaa^{l_a})({\bbb^\dag}^{l_b} \bmaj (l_b)^2 \bbb^{l_b})  \\ 
&\qquad\qquad\qquad\qquad\qquad\qquad\qquad-( \amaj (l_a) \aaa^{l_a}{\aaa^\dag}^{l_a} \amaj (l_a))( \bmaj (l_b) \bbb^{l_b}{\bbb^\dag}^{l_b} \bmaj (l_b))]  \\
&+E_J^2  \sum \limits_{l_a,l_b \geq 1} \frac{((-1)^{l_a}+(-1)^{l_b})}{2(l_a \omega_a-l_b \omega_b)}[({\aaa^\dag}^{l_a} \amaj (l_a)^2 \aaa^{l_a})( \bmaj (l_b) \bbb^{l_b}{\bbb^\dag}^{l_b} \bmaj (l_b))\\
&\qquad\qquad\qquad\qquad\qquad\qquad\qquad-( \amaj (l_a) \aaa^{l_a}{\aaa^\dag}^{l_a} \amaj (l_a))({\bbb^\dag}^{l_b} \bmaj (l_b)^2 \bbb^{l_b})], 
\end{align*}
where $\amaj (l_a)$ and $\bmaj (l_b)$ are introduced in Section~\ref{sec:derivationRWA}. The ratios $\Gamma_\phi^\pm / \Gamma_m$, where $\Gamma_\phi^+$ ($\Gamma_\phi^-$) is the measurement induced dephasing within the even (resp. odd) joint parity subspace, quantify how much the quantum state is disturbed during a parity measurement, and can be evaluated through
\begin{align*}
\Gamma_\phi^+ &=\Big| \bra{\CCC_\alpha^+, \CCC_\beta^+}\hhh^{\text{RWA,2}}\ket{\CCC_\alpha^+,\CCC_\beta^+}-\bra{\CCC_\alpha^-, \CCC_\beta^-}\hhh^{\text{RWA,2}}\ket{\CCC_\alpha^-,\CCC_\beta^-}\Big|/\hbar\notag\\
\Gamma_\phi^- &=\Big|\bra{\CCC_\alpha^+, \CCC_\beta^-}\hhh^{\text{RWA,2}}\ket{\CCC_\alpha^+,\CCC_\beta^-}-\bra{\CCC_\alpha^-, \CCC_\beta^+}\hhh^{\text{RWA,2}}\ket{\CCC_\alpha^-,\CCC_\beta^+}\Big|/\hbar\notag\\
\Gamma_m&=\Big|\bra{\CCC_\alpha^+, \CCC_\beta^+}\hhh^{\text{RWA,2}}\ket{\CCC_\alpha^+,\CCC_\beta^+}+\bra{\CCC_\alpha^-, \CCC_\beta^-}\hhh^{\text{RWA,2}}\ket{\CCC_\alpha^-,\CCC_\beta^-}\notag\\
&\qquad\qquad-\ket{\CCC_\alpha^-,\CCC_\beta^+}\bra{\CCC_\alpha^+, \CCC_\beta^-}\hhh^{\text{RWA,2}}\ket{\CCC_\alpha^+,\CCC_\beta^-}-\bra{\CCC_\alpha^-, \CCC_\beta^+}\hhh^{\text{RWA,2}}\ket{\CCC_\alpha^-,\CCC_\beta^+}\Big|/\hbar. 
\end{align*}
As a numerical example, we set the mode frequencies to $\omega_a = 9.10$GHz and $\omega_b = 7.5$GHz, the Josphson energy $E_J/\hbar=300$MHz, the cat amplitudes $\alpha = \beta = 2$, the parameters $\varphi_a = \varphi_b = 2\alpha$. We find that $\Gamma_\phi^+ / \Gamma_m \sim 10^{-3}$  and $\Gamma_\phi^- / \Gamma_m \sim 5\times 10^{-3}$, while the measurement rate $\Gamma_m$ can be as high as $1$MHz (see Fig. 3 of the main text). 

\section{Zeno dynamics approximation : first order and second order corrections}

\subsection{First order Zeno dynamics approximation}

\subsubsection{Two-photon process : Derivation of the effective hamiltonian ${\hhh^{\text{RWA}}_{\MMM_{2,\alpha}}}$}

Under two-photon driven dissipation, the state of the oscillator is confined to the manifold ${\MMM_{2,\alpha}}= \text{span}\{ \ket{\CCC_\alpha^\pm} \}$. The  dynamics in the first order approximation in $\epsilon_\text{zeno}=E_J/\hbar \kappa_\text{2ph}$, is given by the projection ${\hhh^{\text{RWA}}_{\MMM_{2,\alpha}}}$ of the hamiltonian $ \hhh_\text{RWA}$ on ${\MMM_{2,\alpha}} $. Since $\hhh_\text{RWA}$ is diagonal in the Fock states basis, the off diagonal matrix elements $\bra{\CCC_\alpha^\pm} \hhh_\text{RWA} \ket{\CCC_\alpha^\mp}$ are identically zero. The diagonal matrix elements $c_\alpha^\pm = \bra{\CCC_\alpha^\pm} \hhh_\text{RWA} \ket{\CCC_\alpha^\pm}$ satisfy 
\begin{align}
\label{eq:matrixelementstwophotonsupp}
c_\alpha^\pm &=  -E_J e^{-\frac{\varphi_a^2}{2}-|\alpha|^2} \sum \limits_{n \geq 0} \frac{|\alpha|^{2n}}{n!}L_n(\varphi_a^2) \mp E_J e^{-\frac{\varphi_a^2}{2}-|\alpha|^2} \sum \limits_{n \geq 0} \frac{(-|\alpha|^2)^{n}}{n!}L_n(\varphi_a^2) \notag \\
& = -E_J e^{-\frac{\varphi_a^2}{2}} [ J_0 (2|\alpha| \varphi_a) \pm e^{-2|\alpha|^2}I_0(2|\alpha| \varphi_a)],
\end{align}
where $J_0(.)$ and $I_0(.)$ are respectively the Bessel function and the modified Bessel function, both of the first kind. To derive the second line of eq. (\ref{eq:matrixelementstwophotonsupp}), we applied the identity (5.1.16) of~\cite{OrthogonalPolynomials}. As we have $|J_0(2|\alpha| \varphi_a)| \leq 1$, the first term is bounded by $E_J e^{-\frac{\varphi_a^2}{2}}$. Note that this ensures the symmetry $c_\alpha^+ = - c_\alpha^- + \mathcal{O}(E_J e^{-\frac{\varphi_a^2}{2}})$. Using the asymptotic expansion (29.7) of~\cite{KorenevBesselFunctions} for $I_0$ leads to $e^{-\frac{\varphi_a^2}{2}} e^{-2|\alpha|^2}I_0(2|\alpha| \varphi_a) = e^{-\frac{1}{2}(\varphi_a-2|\alpha|)^2}[1+F(2|\alpha| \varphi_a)]/\sqrt{4 \pi |\alpha| \varphi_a}$, with $F$ satisfying $|F(2|\alpha| \varphi_a)|< (16|\alpha| \varphi_a)^{-1}$. Hence, the hamiltonian ${\hhh^{\text{RWA}}_{\MMM_{2,\alpha}}}$ reads 
\begin{align*} \label{eq:result}
\hhh^{\text{RWA}}_{\MMM_{2,\alpha}} &= -\frac{\hbar\Omega_a}{2}[ \ket{\CCC^{+}_{\alpha}} \bra{\CCC^{+}_{\alpha}}-\ket{\CCC^{-}_{\alpha}}\bra{\CCC^{-}_{\alpha}}] +\mathcal{O}(E_J e^{-\frac{\varphi_a^2}{2}}), \quad \Omega_a = \frac{E_J}{\hbar} \frac{e^{-\frac{1}{2}(\varphi_a-2|\alpha|)^2}}{\sqrt{\pi |\alpha| \varphi_a}}[1+F(2|\alpha| \varphi_a)].  \notag \\
&=  -\frac{\hbar\Omega_a}{2} \sss_z^L+ \mathcal{O}(E_Je^{-\frac{\varphi_a^2}{2}}).
\end{align*}
Note that in the regime we consider, we typically have $\varphi_a = 2 |\alpha| = 4$. It leads to $|F(2|\alpha| \varphi_a)|<  1/64$, so that $\Omega_a$ is well approximated by 
\begin{equation*} 
\Omega_a \approx \frac{E_J}{\hbar} \frac{e^{-\frac{1}{2}(\varphi_a-2|\alpha|)^2}}{\sqrt{\pi |\alpha| \varphi_a}}.
\end{equation*}

\subsubsection{Four-photon process : Derivation of the effective hamiltonian ${\hhh^{\text{RWA}}_{\MMM_{4,\alpha}}}$}
\label{sec:fourphoton}

Under four-photon driven dissipation, the state is confined to the manifold ${\MMM_{4,\alpha}}= \text{span}\{ \ket{\CCC_{\alpha}^{\text{(jmod4)}}},~j=0,1,2,3 \}$, and the effective hamiltonian is given by the projection ${\hhh^{\text{RWA}}_{\MMM_{4,\alpha}}}$ of the physical hamiltonian $\hhh_\text{RWA}$ on the 4D-manifold ${\MMM_{4,\alpha}}$ (first order approximation in $\epsilon_\text{zeno} = E_J /\hbar \kappa_\text{4ph}$). Since $\hhh_\text{RWA}$ is diagonal in the Fock states basis,  its projection ${\hhh^{\text{RWA}}_{\MMM_{4,\alpha}}}$ is also diagonal in the basis $\{ \ket{\CCC_{\alpha}^{\text{(jmod4)}}}\}$ (as the expansion of $ \ket{\CCC_{\alpha}^{\text{(jmod4)}}}$ includes only Fock states $n$ such that $n~mod~4=j$). The diagonal matrix elements $c_\alpha^{jj} = \bra{\CCC_{\alpha}^{\text{(jmod4)}}} \hhh_\text{RWA} \ket{\CCC_{\alpha}^{\text{(jmod4)}}}$ read 
\begin{align*} 
c_\alpha^{jj} = (-1)^{j+1} \frac{\hbar \Omega_a}{2} + \frac{i^j}{2}[\bra{i\alpha} \hhh_\text{RWA} \ket{\alpha} + (-1)^j \bra{i\alpha} \hhh_\text{RWA} \ket{-\alpha} ],\quad j=0,1,2,3
\end{align*}
In the above expression, the first term comes from the coupling of $\ket{\alpha}$ and $\ket{-\alpha}$, and  the coupling of $\ket{i\alpha}$ and $\ket{-i\alpha}$. This corresponds to the desired parity-like term. Note that its amplitude is given by $ {\hbar \Omega_a}/{2}$, as in the case of ${\hhh^{\text{RWA}}_{\MMM_{2,\alpha}}}$. The second term, resulting from the coupling of the states $\ket{\pm\alpha}$ and $\ket{\pm i\alpha}$ through $\hhh_\text{RWA}$, lifts the degeneracy within the parity subspace. This undesired term can be evaluated, as we have
\begin{align*} 
\bra{i\alpha} \hhh_\text{RWA} \ket{\alpha} &= -E_J e^{-\varphi_a^2-(1+i)|\alpha|^2} I_0 (2e^{i\frac{\pi}{4}}|\alpha| \varphi_a) \notag \\
&\approx -E_J e^{-\frac{1}{2}(\varphi_a-\sqrt{2}|\alpha|)^2}\frac{e^{i[-\frac{\pi}{8}+|\alpha|(|\alpha|-\sqrt{2} \varphi_a)]}}{\sqrt{4\pi|\alpha|\varphi_a}},
\end{align*}
and $\bra{i\alpha} \hhh_\text{RWA} \ket{-\alpha} =  (\bra{\alpha} \hhh_\text{RWA} \ket{i\alpha})^\dag$. Hence, the diagonal matrix elements read
\begin{align*} 
& c_\alpha^{jj} = -\frac{\hbar \Omega_a}{2} + (-1)^\frac{j}{2} \frac{e^{-\frac{1}{2}(\varphi_a-\sqrt{2}|\alpha|)^2}}{\sqrt{4\pi|\alpha|\varphi_a}} \cos \big( |\alpha|(|\alpha|-\sqrt{2} \varphi_a)-\frac{\pi}{8} \big),\quad j=0,2 \notag \\
& c_\alpha^{jj} =  \frac{\hbar \Omega_a}{2} + (-1)^\frac{j-1}{2} \frac{e^{-\frac{1}{2}(\varphi_a-\sqrt{2}|\alpha|)^2}}{\sqrt{4\pi|\alpha|\varphi_a}} \sin \big( |\alpha|(|\alpha|-\sqrt{2} \varphi_a)-\frac{\pi}{8} \big),\quad j=1,3.
\end{align*}
Under the condition $\varphi_a = 2|\alpha|$, the second terms scale as $e^{-|(\sqrt{2}-1) \alpha|^2}|\alpha|^{-1}$, and the ratio of the second and the first terms simply scales as $e^{-|(\sqrt{2}-1) \alpha|^2}$. Note that $(\sqrt{2}-1)^2 \approx 0.17$, which explains the necessity to consider larger cat amplitude $\alpha$ to obtain a parity hamiltonian (see Fig. 1d of the main text).

\subsection{Second order Zeno dynamics approximation for two modes under two-photon driven dissipation}
\label{sec:secondorderzeno}

In this section, we study the effect of the second order Zeno dynamics approximation for the case of two modes $\aaa$ and $\bbb$, both subject to two-photon driven dissipation and the two mode hamiltonian $\hhh_\text{RWA}$ given by eq.~\ref{eq:twomodeRWAsupp}, with $\varphi_a \approx 2 |\alpha|$ and $\varphi_b \approx 2 |\beta|$.
While the first order Zeno dynamics approximation in $\epsilon_\text{zeno}$ corresponds to a modification of the hamiltonian part of the dynamics ($\hhh_\text{RWA}$ acts as a projected hamiltonian on the manifold ${\MMM_{2,\alpha,\beta}}$), the second order correction arises in the form of dissipation channels described by Lindblad operators acting on ${\MMM_{2,\alpha,\beta}}$. As the hamiltonian $\hhh_\text{RWA}$ is diagonal in the Fock state basis, these Lindblad operators are linear combinations of $\III_a \otimes \III_b,~ \sss_a \otimes \III_b,~\III_a \otimes \sss_b ~\text{and}~  \sss_a \otimes \sss_b$. Indeed, the hamiltonian $\hhh_\text{RWA}$ cannot induce any change of single-mode parities. It, therefore, cannot induce bit-flip type errors on logical qubits. While the correlated phase-flips occuring through the operator $\sss_a \otimes \sss_b$ reduce the efficiency of the measurement without affecting the state of the system (see main text), independent phase-flips induced by the operators $\sss_a \otimes \III_b$ and $\III_a \otimes \sss_b$ lead to an unwanted dephasing within the parity subspaces. In what follows, we show numerical evidence that the induced dephasing rate $\gamma_\text{ind}$ decreases exponentially with the cats amplitudes $|\alpha|$ and $|\beta|$. For simplicity sakes, we set $\alpha=\beta$. 

It is useful to note that phase-flips in the basis $\{ \ket{\CCC_\alpha^\pm} \}$ corresponds to bit-flips in the basis $\{ \ket{\pm \alpha} \}$. Thus, under independent phase-flips, the state $\ket{\alpha,\alpha}$ will evolve towards a mixture of the states $\ket{\alpha,\alpha}$, $\ket{-\alpha,\alpha}$, $\ket{\alpha,-\alpha}$ and $\ket{-\alpha,-\alpha}$, whereas correlated phase-flips will map $\ket{\alpha,\alpha}$ to a mixture of the states $\ket{\alpha,\alpha}$ and $\ket{-\alpha,-\alpha}$. Initializing the two-mode system in the state $\ket{\alpha,\alpha}$ and letting it evolve for a fixed time $T_\text{2ph}=(\kappa_\text{2ph})^{-1} \gg \gamma_\text{ind}^{-1} $, where $\kappa_\text{2ph} = \min(\kappa^a_\text{2ph},\kappa^b_\text{2ph})$, the quantity $\gamma_\text{ind}/\kappa_\text{2ph}$ is given by to the final state population on the manifold ${\text{span} \{ \ket{-\alpha,\alpha},\ket{\alpha,-\alpha}\}}$. However, simulating the full two-mode master equation for high values of $|\alpha|$, e.g $|\alpha|=5$, requires important computational resources due to the high dimensionality of the Hilbert space. We propose a  different semi-analytical approach to circumvent this difficulty.

The idea is to analytically derive the second order corrective terms acting on the reduced manifold ${\MMM_{2,\alpha,\alpha}}$. As stated in~\cite{Azouit-CDC-2016-sup}, these terms take the form of Lindblad operators $\cmaj_{a,\mu}$ and $\cmaj_{b,\mu}$, given by $\cmaj_{j,\mu} =   \bf{M}_\mu \rmaj_j$, with $\rmaj_j = 2 \lmaj_j (\lmaj_a^\dag \lmaj_a+\lmaj_b^\dag \lmaj_b)^{-1} \hhh_\text{RWA} \boldsymbol{\Pi}_{\MMM_{2,\alpha,\alpha}}$. Here, we have used the operators $\lmaj_a = \sqrt{\kappa_\text{2ph}}(\aaa^2-\alpha^2)$, $\lmaj_b = \sqrt{\kappa_\text{2ph}}(\bbb^2-\alpha^2)$, and the projector $\boldsymbol{\Pi}_{\MMM_{2,\alpha,\alpha}}$ on the manifold ${\MMM_{2,\alpha,\alpha}}$. The inverse operation in $(\lmaj_a^\dag \lmaj_a+\lmaj_b^\dag \lmaj_b)^{-1}$ is understood here as the Moore-Penrose pseudo-inverse. Finally, $\{\bf{M}_\mu\}$ corresponds to a set of Kraus operators such that, for any initial state $\rho$, $\LLL_{2,\alpha,\alpha}(\rho)=\sum_\mu \bf{M}_\mu\rho\bf{M}_\mu^\dag $ is  the steady state of the system  subject to two-photon process. Note that evaluating the application of the Kraus map $\LLL_{2,\alpha,\alpha}$ to a state $\rrr$ only requires the calculation of the nine quantities $\av{{\bf{J}}_{r1,r2}}_{\rrr}=\text{tr}(\rrr {\bf{J}}_{r1,r2}) $, with \small${\bf{J}}_{r1,r2} = {\bf{J}}_{r1}\otimes {\bf{J}}_{r2}$ and $(r1,r2) \in\{ (++,++),(++,--),(--,++),(++,+-),(--,+-),(+-,++),(+-,--),(+-,+-),(+-,-+) \}$\normalsize, and where the single mode operators ${\bf{J}}_{++}$ and ${\bf{J}}_{+-}$ are defined in eq. (A.1) and (A.2) of~\cite{Mirrahimi-al-NJP-2014-sup}. 

\begin{figure}[h]
  \centering 
  \hspace*{-1.4cm} 
{\includegraphics[width=0.45\textwidth]{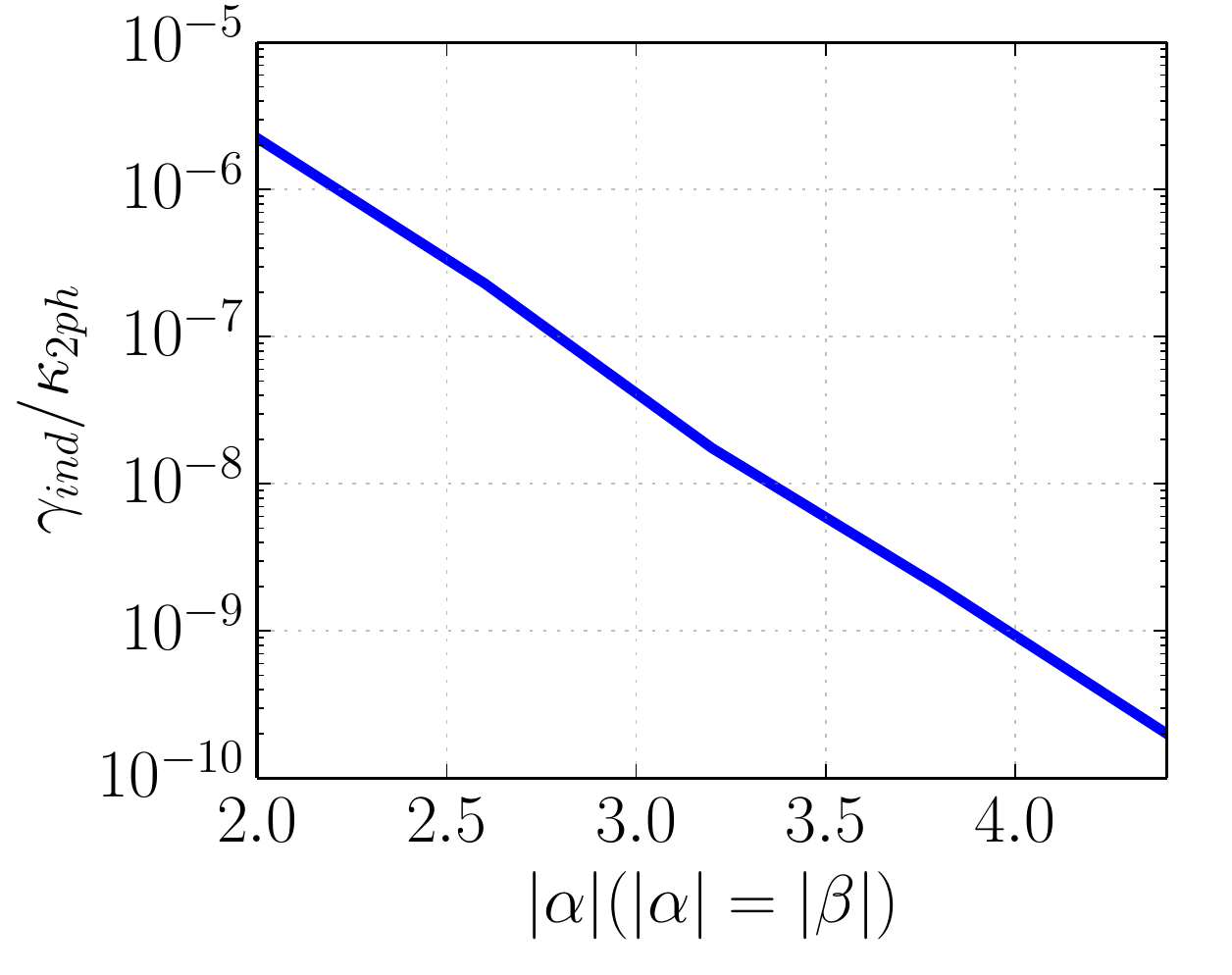}}
  \caption{Numerical estimation of undesired dephasing $\gamma_\text{ind}$ (see eq. (\ref{eq:gammaind_expsupp})) within the two-mode parity subspaces. This dephasing rate,  induced by second order correction to Zeno dynamics, is exponentially suppressed with $|\alpha|$. \label{fig:gammaind}}
\end{figure}

In the second order Zeno dynamics approximation, the reduced master equation reads\small
\begin{align}
\frac{d \rrr_\text{2nd}}{dt} =& -i[\hhh_{\MMM_{2,\alpha,\alpha}}/\hbar,\rrr_\text{2nd}]+\sum \limits_{\mu; j=a,b} \bf{M}_\mu \rmaj_j \rrr_\text{2nd} \rmaj_j^\dag \bf{M}_\mu^\dag -\frac{1}{2} (\rmaj_j^\dag  \bf{M}_\mu^\dag \bf{M}_\mu \rmaj_j \rrr_\text{2nd} + \rrr_\text{2nd} \rmaj_j^\dag  \bf{M}_\mu^\dag \bf{M}_\mu \rmaj_j ) \notag \\
=& -i[\hhh_{\MMM_{2,\alpha,\alpha}}/\hbar,\rrr_\text{2nd}]+ \sum \limits_{j=a,b} \LLL_{\MMM_{2,\alpha,\alpha}}(\rmaj_j \rrr_\text{2nd} \rmaj_j^\dag)  -\frac{1}{2} (\rmaj_j^\dag \rmaj_j \rrr_\text{2nd} + \rrr_\text{2nd} \rmaj_j^\dag \rmaj_j ) \\
\end{align}\normalsize
Here, we study the second order correction 
$${\mathcal{R}_{\text{2nd}} (\rrr_\text{2nd}) = \sum \limits_{j=a,b} \LLL_{\MMM_{2,\alpha,\alpha}}(\rmaj_j \rrr_\text{2nd} \rmaj_j^\dag)  -\frac{1}{2} (\rmaj_j^\dag \rmaj_j \rrr_\text{2nd} + \rrr_\text{2nd} \rmaj_j^\dag \rmaj_j )}.$$ 
The rate $\gamma_\text{ind}$ is well estimated by 
\begin{align} \label{eq:gammaind_expsupp}
\gamma_\text{ind} %\sim \bra{-\alpha,\alpha} \bf{D}_{\text{2nd}}(\ket{\alpha,\alpha}\bra{\alpha,\alpha}) \ket{-\alpha,\alpha}+\bra{\alpha,-\alpha} \bf{D}_{\text{2nd}} (\ket{\alpha,\alpha}\bra{\alpha,\alpha}) \ket{\alpha,-\alpha}
\sim \text{tr}[(\ket{-\alpha,\alpha}\bra{-\alpha,\alpha}+\ket{\alpha,-\alpha}\bra{\alpha,-\alpha})\mathcal{R}_{\text{2nd}} (\ket{\alpha,\alpha}\bra{\alpha,\alpha})].
%\sum \limits_{j=a,b}\bra{-\alpha,\alpha} \LLL_{\MMM_{2,\alpha,\alpha}}(\rmaj_j \ket{\alpha,\alpha}\bra{\alpha,\alpha} \rmaj_j^\dag)\ket{-\alpha,\alpha} +\bra{\alpha,-\alpha} \LLL_{\MMM_{2,\alpha,\alpha}}(\rmaj_j \ket{\alpha,\alpha}\bra{\alpha,\alpha} \rmaj_j^\dag)\ket{\alpha,-\alpha}%= \sum \limits_{j=a,b} \text{tr} \big[ \big(\frac{1-\bf{J}_X^a \bf{J}_X^b}{2}\big) \rmaj_j \ket{\alpha,\alpha}\bra{\alpha,\alpha} \rmaj_j^\dag \big],
\end{align}
In Fig.~\ref{fig:gammaind}, we numerically calculate the above expression. We clearly observe an exponential suppression of $\gamma_\text{ind}$ with the cats amplitude $|\alpha|$, for $\epsilon_\text{zeno}=1$.

This exponential suppression can be understood through the study of the jump operators $\rmaj_j$. We first focus on a single mode case, where the dissipation part is given  ${\mathcal{R}_{\text{2nd}}(\rrr_\text{2nd}) =  \LLL_{\MMM_{2,\alpha}}(\rmaj \rrr_\text{2nd} \rmaj^\dag)  -\frac{1}{2} (\rmaj^\dag \rmaj \rrr_\text{2nd} + \rrr_\text{2nd} \rmaj^\dag \rmaj )}$, $\rmaj = 2 \lmaj_a (\lmaj_a^\dag \lmaj_a)^{-1} \hhh_\text{RWA} \boldsymbol{\Pi}_{\MMM_{2,\alpha,\alpha}}$ and $\hhh_\text{RWA}$ is given in eq.~(\ref{eq:onemodeRWAsupp}). The state $\ket{\alpha}$ is mapped through a jump to the state $\ket{\psi_R}=\rmaj \ket{\alpha}/||\rmaj \ket{\alpha}||$. The first row of Fig.~\ref{fig:wig} shows the Husimi Q functions $Q(\gamma)= \frac{1}{\pi}|\bra{\gamma} \psi_R \rangle |^2$ for $\alpha = 2,3,4,5$. While the states population is concentrated around $\ket{-\alpha}$, it presents a dip precisely at ${-\alpha}$. This dip simply indicates that the dissipation term arises from the component of $\hhh_\text{RWA}$ which differs from the parity hamiltonian. The Q functions of the projected states $\LLL_{\MMM_{2,\alpha}}(\ket{\psi_R}\bra{\psi_R})$, represented in second row of Fig.~\ref{fig:wig}, show a quick suppression of the final population on the state $\ket{\alpha}$. This means that the second order Zeno effect only leads to a $\sss_z$  jump and not a combination of $\III$ and $\sss_z$. Similarly, in the two-mode case, the jumps are necessarily of the form $\sss_z^a\otimes \sss_z^b$.

%some states $\ket{\beta}$ with $Re(\beta) \approx 0$ are also populated for small $|\alpha|$, e.g $\alpha=2$. The latter states $\ket{\beta}$ with $Re(\beta) \approx 0$ are mapped to an incoherent mixture of $\ket{-\alpha}$ and $\ket{\alpha}$ through the projective application $\LLL_{\MMM_{2,\alpha}}$ (see~\ref{Mirrahimi-al-NJP-2014}). The population on $\ket{\alpha}$, which is responsible for the independent phase-flip in the two mode case (through the creation of the states $\ket{\alpha,-\beta}$), seems to decrease exponentially with the cat amplitude $|\alpha|$. 
%In the two mode case, the effect of the dissipation operators $\rmaj_j$ is very similar to that of $\rmaj$. Hence, we have  $\NNN_1 \sum \limits_{j=a,b} \LLL_{\MMM_{2,\alpha,\beta}}(\rmaj_j \rrr_\text{2nd} \rmaj_j^\dag) = \ket{-\alpha,-\beta}\bra{-\alpha,-\beta}) +S(|\alpha|)) $, where $S(|\alpha|)$ seems to decrease exponentially in $|\alpha|$. 

%\begin{figure}[h]
%  \centering
%    \centering
%      \subfloat[]{\includegraphics[width=0.5\textwidth]{Figures/Jxvaryalpha.pdf}\label{fig:Jxvaryalpha}}
%   \hfill
%  \subfloat[]{\includegraphics[width=0.5\textwidth]{Figures/gammaind.pdf}\label{fig:gammaind}}
%  \hfill
%  \caption{(a) :  (b) : .}
%\end{figure}

\begin{figure}[htp!]
  \centering \label{fig:wig}
{\includegraphics[width=\textwidth]{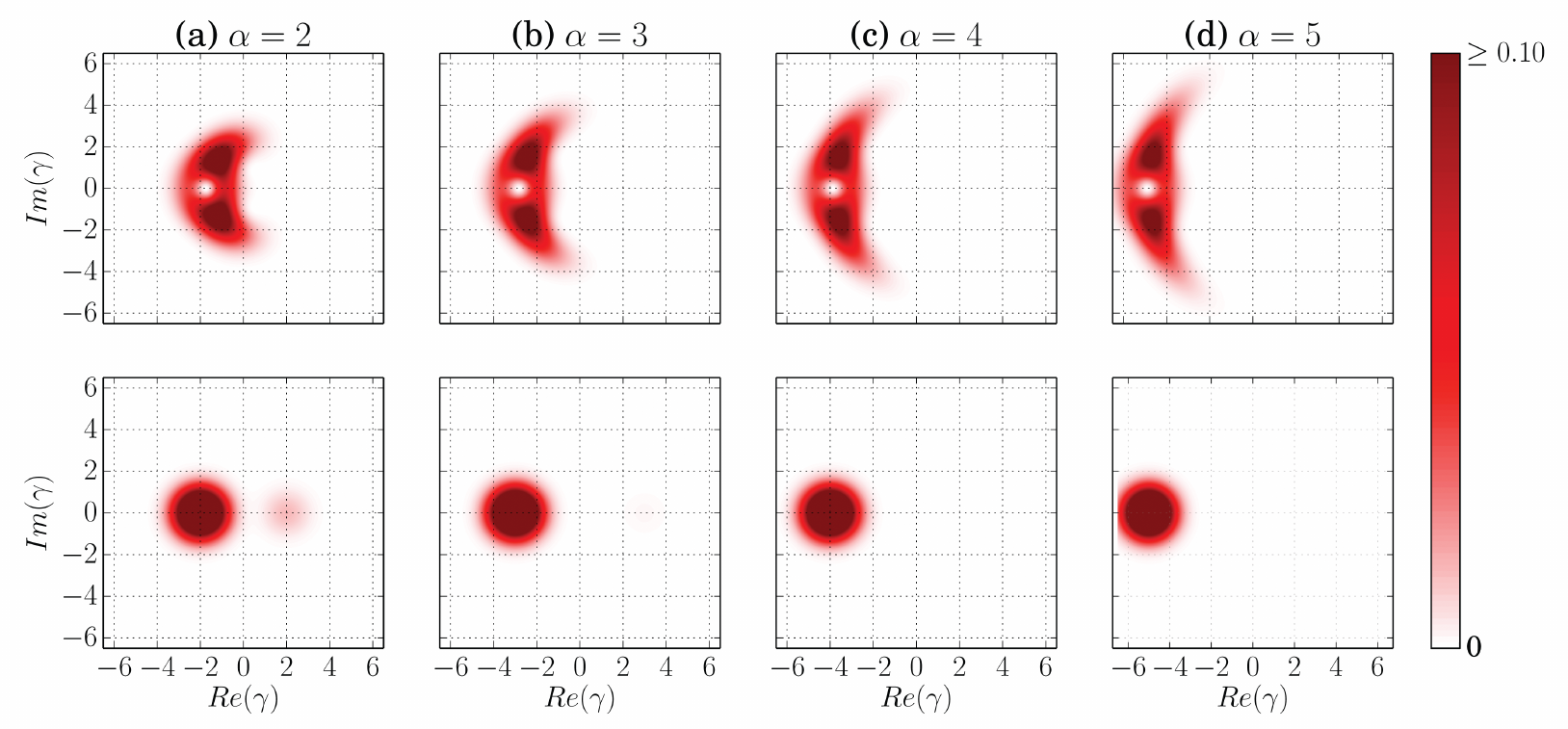}}
 \caption{The first row shows the Husimi Q functions $Q(\gamma)= \frac{1}{\pi}|\bra{\gamma} \psi_R \rangle |^2$ of the state $\ket{\psi_R}=\rmaj \ket{\alpha}/||\rmaj \ket{\alpha}||$, where $\rmaj$ is the single mode jump operator defined in Section~\ref{sec:secondorderzeno}. The different figures correspond to the various values of $\alpha = 2,3,4,5$, from left to right. While the states population is concentrated around $\ket{-\alpha}$, it presents a dip precisely at ${-\alpha}$. This dip simply indicates that the dissipation term arises from the component of $\hhh_\text{RWA}$ which differs from the parity hamiltonian. For the same values of $\alpha$, the second row shows the Q functions of the projected states $\LLL_{\MMM_{2,\alpha}}(\ket{\psi_R}\bra{\psi_R})$. We observe a rapid suppression of the population on $\ket{\alpha}$ with the amplitude $\alpha$. As numerically estimated in Fig.~\ref{fig:gammaind}, this suppression is exponential. \label{fig:wig}}
\end{figure}

\section{Perfect degeneracy of the parity hamiltonian under four photon loss using three junctions}

\begin{figure}[htp!]
  \centering
{\includegraphics[width=0.35\textwidth]{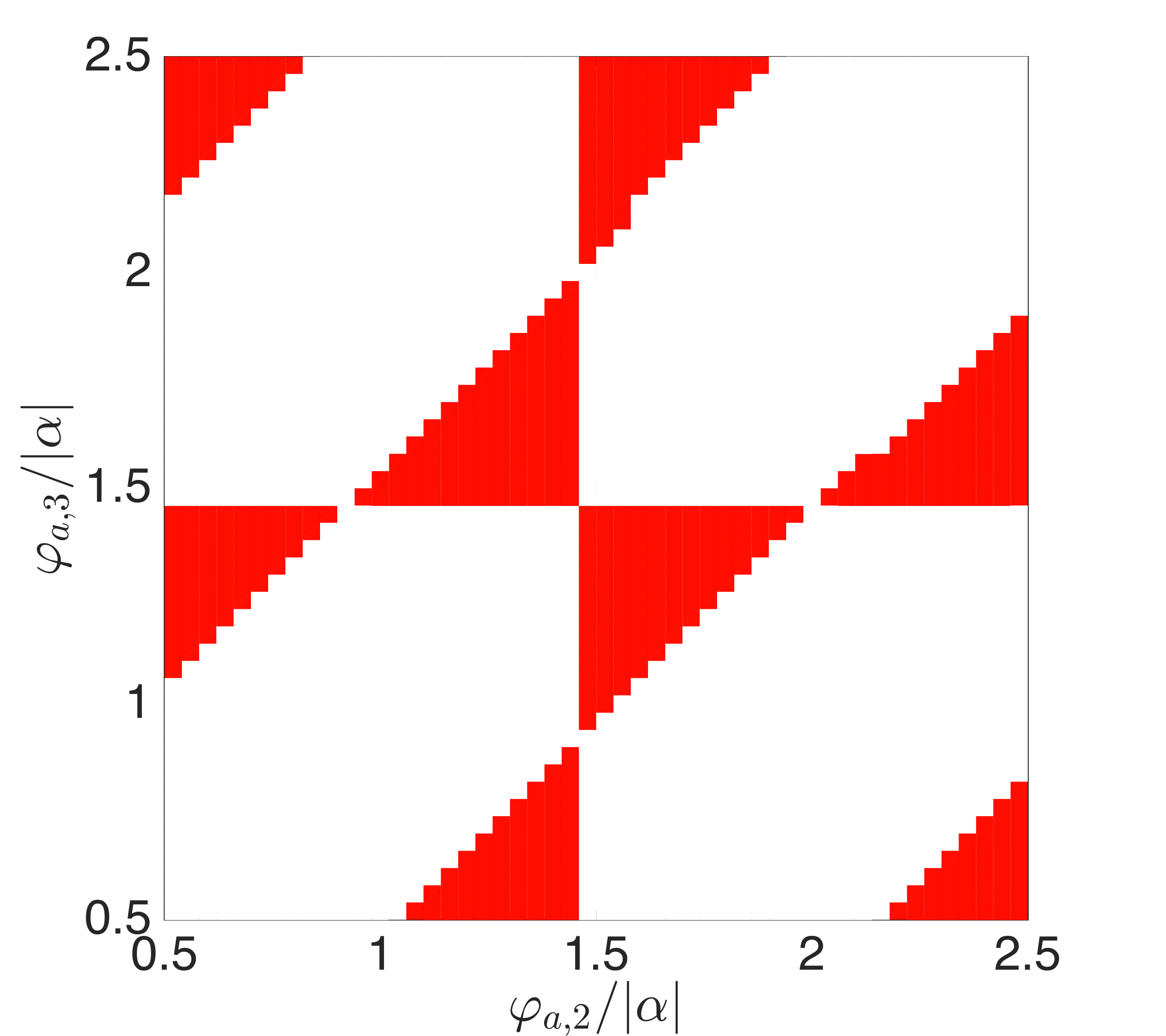}}
  \caption{By choosing $\varphi_{a,1}=2|\alpha|$ and $|\alpha|=2$, we indicate in red the domains of $\varphi_{a,2}$ and $\varphi_{a,3}$ (in units of $|\alpha|$) for which the hamiltonian $\hhh^\text{RWA}$, given by~\ref{eq:threejunchamiltoniansupp}, can potentially act as a perfect parity hamiltonian on $\MMM_{4,\alpha}$. More precisely, one can carefully choose the Josephson energies $E_{J,1}$, $E_{J,2}$ and $E_{J,3}$, such that the parity subspaces of $\hhh_{\MMM_{4,\alpha}}^\text{RWA}$ are degenerate. \label{fig:threejunc} }
\end{figure}

In the Letter, it was pointed out that the projection of the hamiltonian $\hhh^{\text{RWA},1}$ on the 4D-subspace $\MMM_{\alpha,4}$ does not lead to an exact parity hamiltonian. More precisely, we have ${\hhh^{\text{RWA}}_{\MMM_{4,\alpha}} \propto  \Pi_{\MMM_{4,\alpha}}\cos(\pi\aaa^\dag\aaa)\Pi_{\MMM_{4,\alpha}} + O(e^{-\xi |\alpha|^2})}$. While ${\hhh^{\text{RWA}}_{\MMM_{4,\alpha}}}$ approaches the parity hamiltonian exponentially with the cat size $|\alpha|^2$, the parity subspace are non-degenerate for small $|\alpha|$ (see Section~\ref{sec:fourphoton}). This non-degeneracy is illustrated in Figure 1c of the manuscript. Here, we show that one can make the parity subspace exactly degenerate by capacitively coupling a high impedance cavity mode to three Josephson junctions instead of a single one. Considering the hamiltonian of such a system, we move to the rotating frame and apply the RWA, leading to
\begin{align} \label{eq:threejunchamiltoniansupp}
\hhh^{\text{RWA}} = \hhh_1^{\text{RWA}}+\hhh_2^{\text{RWA}}+\hhh_3^{\text{RWA}},\quad  \hhh_k^{\text{RWA}} = -E_{J,k} \sum \limits_n e^{-\frac{\varphi_{a,k}^2}{2}} L_{n}(\varphi_{a,k}^2) \ket{n}\bra{n}. 
\end{align}
$E_{J,i}$ is the Josephson energy of junction $i$, and $\varphi_{a,i}= \sqrt{{\pi e^2 Z_{a,i}}/{h}}$ where $Z_{a,i}$ is the impedance of the cavity mode seen by junction $i$. We define $c_{k}^{jj} = \bra{\CCC_\alpha^{(j mod 4)}} \hhh_k^{\text{RWA}} \ket{\CCC_\alpha^{(j mod 4)}}$, $j=0,1,2,3$, and  $k=1,2,3$. We then note $\Delta^{0,2}_k = c^{22}_k - c^{00}_k$ and $\Delta^{3,1}_k = c^{11}_k - c^{33}_k$, $k=1,2,3$. The parity subspaces are degenerate if one has
\begin{align}
& \sum \limits_{k=1,2,3} E_{J,k} \Delta^{0,2}_k = 0 \notag \\
& \sum \limits_{k=1,2,3} E_{J,k} \Delta^{3,1}_k = 0 \notag  \\
& \quad E_{J,k}>0.
\end{align}
We think of the Josephson energies $E_{J,k}$ as the variables of the system as they can be effectively adjusted using SQUID architecture, while the $\varphi_{a,k}$ are parameters of the system. 
The system only needs a rough tuning of the parameters $\varphi_{a,k}$. In Fig.~\ref{fig:threejunc}, the red domains indicate the existence of a solution $(E_{J,1},E_{J,2},E_{J,3})$ for the above system by as we vary $\varphi_{a,2}$ and $\varphi_{a,3}$ from $0.5 |\alpha|$ to $2.5 |\alpha|$, having set $\varphi_{a,1}=2|\alpha|$ and $|\alpha|=2$. The system requires a rough tuning of $\varphi_{a,2}$ and $\varphi_{a,3}$ and a fine tuning of the Josephson energies to achieve perfect degeneracy of the parity subspaces.

\section{Extension : from a $\mathbb{Z}_2$-parity hamiltonian to a $\mathbb{Z}_n$-parity hamiltonian}
\begin{figure}[h]
  \centering
    \centering
      \subfloat[]{\includegraphics[width=0.5\textwidth]{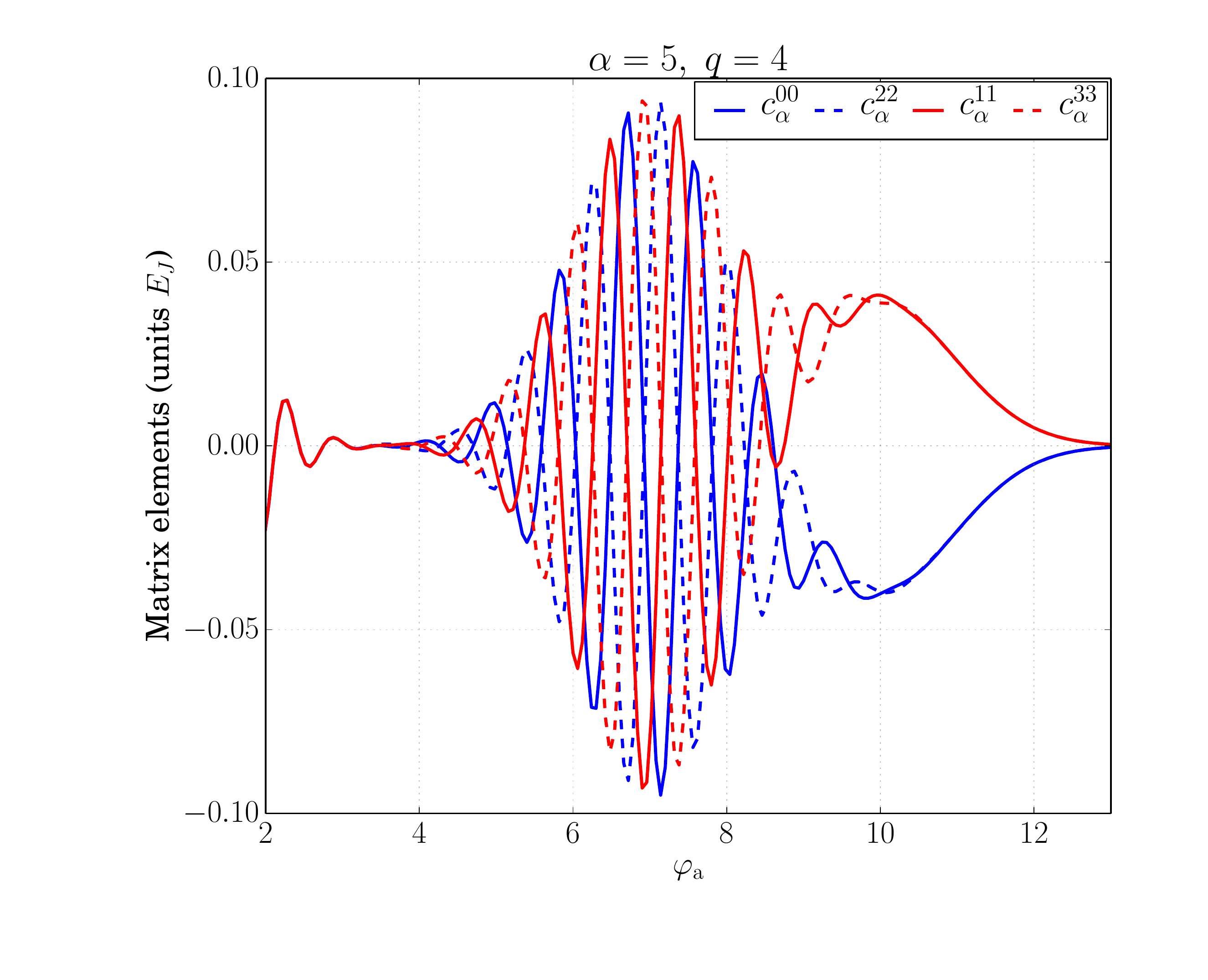}\label{fig:matrixelements4photonalpha5_sectionsupp}}
   \hfill
  \subfloat[]{\includegraphics[width=0.5\textwidth]{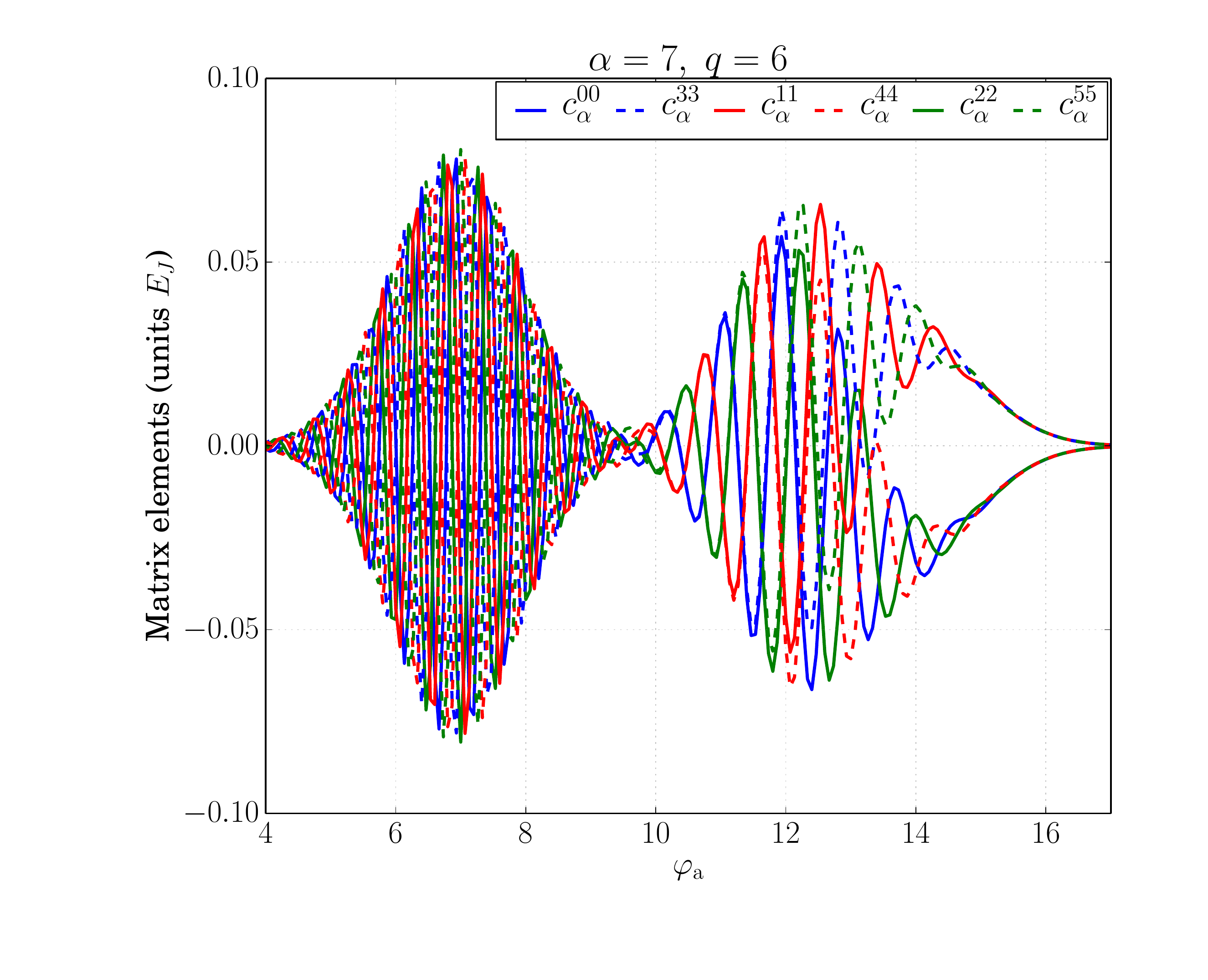}\label{fig:matrixelements6photonalpha7_sectionsupp}}
  \hfill
    \caption{(a) : Diagonal matrix elements $c^{kk}_{\alpha=5,q=4}=\bra{\CCC_{\alpha=5}^{(k \mod 4)}} \hhh^\text{RWA} \ket{\CCC_{\alpha=5}^{(k \mod 4)}}$ as given by eq. (\ref{eq:matrixgeneralcasesupp}), plotted as a function of $\varphi_a$. For $\varphi_a \approx 2|\alpha|$ which corresponds to $m_0 = 2$ in eq. (\ref{eq:generalcasesupp}), one has $c^{00}_{\alpha,4} = c^{22}_{\alpha,4}=-c^{11}_{\alpha,4}=-c^{33}_{\alpha,4}$, so that $\hhh^\text{RWA}$ acts as a $\mathbb{Z}_2$-parity hamiltonian on ${\MMM_{4,\alpha}}$. For $\varphi_a\approx \sqrt{2} |\alpha|$ ($m_0=1$), the hamiltonian $\hhh_{\MMM_{4,\alpha}}$ is fully non-degenerate for most values of $\varphi_a$ around $\sqrt{2} |\alpha|$, which allows the measurement of the photon number modulo 4.  (b) : Diagonal matrix elements $c^{kk}_{\alpha=7,q=6} = \bra{\CCC_{\alpha=7}^{(k \mod 6)}} \hhh^\text{RWA} \ket{\CCC_{\alpha=7}^{(k \mod 6)}}$ as given by eq. (\ref{eq:matrixgeneralcasesupp}), plotted as a function of $\varphi_a$. For $\varphi_a \approx 2|\alpha|$ (corresponds to $m_0 = 3$ in eq. (\ref{eq:generalcasesupp}) ),  $\hhh^\text{RWA}$ acts almost as a $\mathbb{Z}_2$-parity hamiltonian on ${\MMM_{6,\alpha}}$. The degeneracy within the parity subspaces is lifted because of the second term of eq. (\ref{eq:generalcasesupp}). For $\varphi_a \approx \sqrt{3} |\alpha|$ ($m_0 = 2$), one has $c^{00}_{\alpha,6} = c^{33}_{\alpha,6}$, $c^{11}_{\alpha,6}=c^{44}_{\alpha,6}$ and $c^{11}_{\alpha,6}=c^{44}_{\alpha,6}$, meaning that the hamiltonian distinguish the states according to their photon number modulo 3. For $\varphi_a \approx |\alpha|$ ($m_0 = 1$), the hamiltonian $\hhh_{\MMM_{6,\alpha}}$ is fully non-degenerate except for a discrete set of values of $\varphi_a$, and allows the measurement of the photon number modulo 6. }
\end{figure}

We have focused on obtaining a $\mathbb{Z}_2$-parity hamiltonian under two-photon (or four-photon) driven dissipation, i.e a physical hamiltonian $\hhh$ satisfying $\hhh_{\MMM_{2(4),\alpha}} \propto \proj_{\MMM_{2(4),\alpha}} \cos(\pi \aaa^\dag \aaa) \proj_{\MMM_{2(4),\alpha}}$. More precisely, under two-photon process, we have designed schemes to measure, in a continuous and QND manner, the observable $\sss_Z^a=\ket{\CCC_\alpha^+}\bra{\CCC_\alpha^+}-\ket{\CCC_\alpha^-}\bra{\CCC_\alpha^-}$ for a single-mode under two-photon process. Under four-photon driven dissipation,  we have proposed a measurement scheme for the observable $\boldsymbol{\pi}_{4\text{ph}}=\ket{\CCC_\alpha^{(0\text{mod}4)}}\bra{\CCC_\alpha^{(0\text{mod}4)}}+\ket{\CCC_\alpha^{(2\text{mod}4)}}\bra{\CCC_\alpha^{(2\text{mod}4)}}-\ket{\CCC_\alpha^{(1\text{mod}4)}}\bra{\CCC_\alpha^{(1\text{mod}4)}}-\ket{\CCC_\alpha^{(3\text{mod}4)}}\bra{\CCC_\alpha^{(3\text{mod}4)}}$.   More generally, under the $q$-photon driven dissipation induced by the Lindblad operator $\sqrt{\kappa_q}(\aaa^q-\alpha^q)$, the oscillator state is confined to the manifold $\MMM_{q,\alpha}=\text{span}  \{ \ket{\alpha e^{\frac{2i\pi p}{q}}},~ p=0,...,q-1 \}$. Considering the same single mode hamiltonian $\hhh^\text{RWA}$ given in eq. (\ref{eq:onemodeRWAsupp}), we show that $\hhh^\text{RWA}$ can act as the operator $\cos (\frac{m\pi}{q}\aaa^\dag \aaa)$ for an appropriate choice of $\varphi_a$.   This operator with $0 \leq m \leq q-1$ corresponds to generalized parity-type observables on $\MMM_{q,\alpha}$.

Let us first define another basis of the manifold $\MMM_{q,\alpha}$, the set of the q-component cat states $\{ \ket{\CCC_\alpha^{(k\mod q)}} = \frac{1}{\sqrt{q}} \sum \limits_{p=0}^{q-1} e^{-\frac{2ipk\pi}{q}} \ket{\alpha e^{\frac{2ip \pi}{q}}} , ~k=0,...,q-1 \}$~\cite{haroche-raimond:book06}. Note that the expansion of the state $\ket{\CCC_\alpha^{(k \mod q)}}$ contains only Fock states $\ket{m}$ such that $m\equiv k (\text{mod}~q)$. With similar arguments used to derive $\hhh^\text{RWA}_{\MMM_{2(4),\alpha}}$, the single mode hamiltonian $\hhh^\text{RWA}$ given by eq. (\ref{eq:onemodeRWAsupp}) acts as the projected hamiltonian on $\MMM_{q,\alpha}$ and its projection is diagonal in the cat states basis. For any $\varphi_a >0 $, the matrix elements $c^{kk}_{\alpha,q} =\bra{\CCC_\alpha^{(k \mod q)}} \hhh^\text{RWA} \ket{\CCC_\alpha^{(k \mod q)}} $ read
\begin{align}\label{eq:matrixgeneralcasesupp}
c^{kk}_{\alpha,q} = -\frac{E_J}{\sqrt{4 \pi \varphi_a |\alpha|}}\sum \limits_{m=0}^{q-1} e^{-\frac{1}{2}(\varphi_a-2|\alpha| \sin(\frac{m \pi}{q}))^2} \cos(k\frac{m\pi}{q}+\theta_{m,q}),
\end{align}
where $\theta_{m,q} = 2|\alpha|\cos(\frac{m \pi}{q})[\varphi_a-|\alpha|\sin(\frac{m\pi}{q})]-\frac{\pi}{4}-\frac{m \pi}{2 q}$. Thus, if one sets $\varphi_a = 2|\alpha| \sin(\frac{m_0 \pi}{q})$, the matrix elements $c^{kk}_{\alpha,q}$ read 
\begin{align} \label{eq:generalcasesupp}
c^{kk}_{\alpha,q} = -\frac{E_J}{\sqrt{4 \pi \varphi_a |\alpha|}} [\cos(k\frac{m_0\pi}{q}+\theta_{m_0,q})+\mathcal{O}(e^{-2|\alpha|^2(\sin \frac{(m_0+1)\pi}{q}-\sin \frac{m_0\pi}{q})^2})],
\end{align}
with $\theta_{m_0,q} = |\alpha^2| \sin(\frac{2 m_0 \pi}{q})-\frac{\pi}{4}-\frac{m_0 \pi}{2 q}$. 

In Fig.~\ref{fig:matrixelements4photonalpha5_sectionsupp}, the matrix elements $c^{kk}_{\alpha,q}$ are represented as a function of $\varphi_a$ in the case $\alpha=5$ and $q=4$ (Fig.~\ref{fig:matrixelements4photonalpha5_sectionsupp}). Around the values $\varphi_a = 2|\alpha| \sin(\frac{m_0 \pi}{4})$, the spectrum exhibits the expected behaviours. For $m_0 = 2$ ($\varphi_a =2|\alpha|$), the spectrum of the hamiltonian $\hhh_{\MMM_{4,\alpha}}$ consists in two degenerate eigenspaces, grouping the states by their photon number modulo 2. For $m_0 = 1$ ($\varphi_a =\sqrt{2}|\alpha|$), the spectrum of the hamiltonian $\hhh_{\MMM_{4,\alpha}}$  is fully non-degenerate, resulting in a distinction of the states according to their photon number modulo 4. In the case $\alpha=7$ and $q=6$ (Fig.~\ref{fig:matrixelements6photonalpha7_sectionsupp}), the stabilized manifold is generated by the states $\ket{\CCC_\alpha^{(k \mod 6)}},~k=0,...,5$ . Setting the parameter $m_0 = 3$ ($\varphi_a =2|\alpha|$), $\hhh_{\MMM_{4,\alpha}}$ gives a parity hamiltonian (photon number modulo 2). For $m_0 = 2$ ($\varphi_a = \sqrt{3} |\alpha|$), there are three degenerate eigenspaces representing the states photon number modulo 3. Finally, the case $m_0 = 1$ ($\varphi_a = |\alpha|$) yields a fully non-degenerate hamiltonian, which allows a spectral distinction of the states indexed by the photon number modulo 6.

\end{document}